\documentclass[twocolumn]{aastex63} %

\newcommand{\fc}{1297.5} % central frequency (MHz)
\newcommand{\bw}{336} % bandwidth (MHz)
\newcommand{\sntab}{220} % S/N when coherently beamformed (226.7 when measured by Hyerin-made software at the highest time resolution)
\newcommand{\snics}{19.3} % S/N when incoherently beamformed

\newcommand{\mdmunits}{{\rm pc \, cm^{-3}}} 

\newcommand{\dmval}{589.265}   % DM in pc/cm^3
\newcommand{\dmerr}{0.001}   % sigma(DM) in pc/cm^3
\newcommand{\trddmval}{589.26}   % third peak's DM in pc/cm^3
\newcommand{\trddmerr}{0.01}   % third peak's sigma(DM) in pc/cm^3
\newcommand{\fthdmval}{589.306}   % fourth peak's DM in pc/cm^3
\newcommand{\fthdmerr}{0.004}   % fourth peak's sigma(DM) in pc/cm^3

\newcommand{\mrmunits}{{\rm rad \, m^{-2}}}

\newcommand{\rmval}{10.5} %{11.8}
\newcommand{\rmerr}{4} %{0.8}
\newcommand{\trdrmval}{25}
\newcommand{\trdrmerr}{2}

\newcommand{\rndrmval}{10.5} %{11.8} % rounded RM value (1st pulse)
\newcommand{\rndrmerr}{0.4} %{0.8} % rounded RM error
\newcommand{\rndtrdrmval}{25} %{26} % rounded RM value (3rd pulse)
\newcommand{\rndtrdrmerr}{2} %{4} % rounded RM error

\newcommand{\mdtpulse}{\tau_{\rm scatt}}
\newcommand{\tscatt}{25} % lower limit of scattering timescale in microsec. (19.12.08. excluding upper 8 MHz)

\newcommand{\freqdc}{4.2} % 19.12.08. excluding upper 8 MHz

\graphicspath{{./}{figures/}}

\usepackage[]{verbatim}

\usepackage{amsmath}
\usepackage{pgf}

\usepackage{pifont}
\usepackage{upgreek}

\newcommand{\xmark}{\text{\ding{55}}}
\renewcommand{\mu}{\upmu} 

\newcommand{\Poincare}{Poincar\'e~}

\received{December 21, 2019}
\revised{February 18, 2020}
\accepted{February 19, 2020} %\today
\submitjournal{ApJL}

\shorttitle{FRB\,181112 at high time resolution}
\shortauthors{Cho et al.}
\begin{document}

\title{Spectropolarimetric analysis of FRB\,181112 at microsecond resolution: Implications for Fast Radio Burst emission mechanism}

\correspondingauthor{Hyerin Cho}
\email{chyerin1996@gmail.com}

\author[0000-0002-2858-9481]{Hyerin Cho}
\affiliation{School of Physics and Chemistry, Gwangju Institute of Science and Technology, Gwangju, 61005, Korea}
\author{Jean-Pierre Macquart}
\affiliation{International Centre for Radio Astronomy Research, Curtin Institute of Radio Astronomy, Curtin University, Perth, WA 6845, Australia}
\author{Ryan~M.~Shannon}
\affiliation{Centre for Astrophysics and Supercomputing, Swinburne University of Technology, Hawthorn VIC 3122, Australia}
\author{Adam T. Deller}
\affiliation{Centre for Astrophysics and Supercomputing, Swinburne University of Technology, Hawthorn VIC 3122, Australia}
\author{Ian S. Morrison}
\affiliation{International Centre for Radio Astronomy Research, Curtin Institute of Radio Astronomy, Curtin University, Perth, WA 6845, Australia}
\author{Ron D. Ekers}
\affiliation{CSIRO Astronomy and Space Science, PO Box 76, Epping, NSW 1710, Australia}
\affiliation{International Centre for Radio Astronomy Research, Curtin Institute of Radio Astronomy, Curtin University, Perth, WA 6845, Australia}
\author{Keith W. Bannister}
\affiliation{CSIRO Astronomy and Space Science, PO Box 76, Epping, NSW 1710, Australia}
\author{Wael Farah}
\affiliation{Centre for Astrophysics and Supercomputing, Swinburne University of Technology, Hawthorn VIC 3122, Australia}
\author{Hao Qiu}
\affiliation{Sydney Institute for Astronomy, School of Physics, University of Sydney, NSW 2006, Australia}
\affiliation{CSIRO Astronomy and Space Science, PO Box 76, Epping, NSW 1710, Australia}
\author{Mawson W. Sammons}
\affiliation{International Centre for Radio Astronomy Research, Curtin Institute of Radio Astronomy, Curtin University, Perth, WA 6845, Australia}
\author{Matthew Bailes}
\affiliation{Centre for Astrophysics and Supercomputing, Swinburne University of Technology, Hawthorn VIC 3122, Australia}
\author{Shivani Bhandari}
\affiliation{CSIRO Astronomy and Space Science, PO Box 76, Epping, NSW 1710, Australia}
\author{Cherie K. Day}
\affiliation{Centre for Astrophysics and Supercomputing, Swinburne University of Technology, Hawthorn VIC 3122, Australia}
\affiliation{CSIRO Astronomy and Space Science, PO Box 76, Epping, NSW 1710, Australia}
\author{Clancy~W.~James}
\affiliation{International Centre for Radio Astronomy Research, Curtin Institute of Radio Astronomy, Curtin University, Perth, WA 6845, Australia}
\author{Chris J. Phillips}
\affiliation{CSIRO Astronomy and Space Science, PO Box 76, Epping, NSW 1710, Australia}
\author{J. Xavier Prochaska}
\affiliation{University of California, Santa Cruz, 1156 High St., Santa Cruz, CA 95064, USA}
\affiliation{Kavli Institute for the Physics and Mathematics of the Universe (Kavli IPMU), 5-1-5 Kashiwanoha, Kashiwa, 277-8583, Japan}
\author{John Tuthill}
\affiliation{CSIRO Astronomy and Space Science, PO Box 76, Epping, NSW 1710, Australia}
%\collaboration{1}{(CRAFT collaboration)}

%% Note that the \and command from previous versions of AASTeX is now
%% depreciated in this version as it is no longer necessary. AASTeX 
%% automatically takes care of all commas and "and"s between authors names.

%% AASTeX 6.3 has the new \collaboration and \nocollaboration commands to
%% provide the collaboration status of a group of authors. These commands 
%% can be used either before or after the list of corresponding authors. The
%% argument for \collaboration is the collaboration identifier. Authors are
%% encouraged to surround collaboration identifiers with ()s. The 
%% \nocollaboration command takes no argument and exists to indicate that
%% the nearby authors are not part of surrounding collaborations.

%% Mark off the abstract in the ``abstract'' environment. 
\begin{abstract}
% RMS:  check abstract length requirements (HC: should be < 250 words. We need to cut it down)
% RMS: worth trying to write the abstract in a more general way

%Understanding the production of fast radio burst emission is one of the major open questions in astrophysics.
%Despite operating on comparable (millisecond) timescales to pulses from Galactic radio pulsars and magnetars, fast radio bursts require an emission process capable of generating bursts twelve orders of magnitude more powerful.
%To date, much theoretical work has focused on the first known repeating fast radio burst, FRB 121102, where dedicated observations have captured detailed information on the radio properties and the host galaxy environment. 
%Comprehensive studies of the emission characteristics of the larger population of one-off bursts have been hindered by insufficient time resolution to resolve burst temporal structure and lack of full polarimetric information.
We have developed a new coherent dedispersion mode to study the emission of Fast Radio Bursts that trigger the voltage capture capability of the Australian SKA Pathfinder (ASKAP) interferometer. In principle the mode can probe emission timescales down to 3\,ns with full polarimetric information preserved.
Enabled by the new capability, here we present a spectropolarimetric analysis of FRB\,181112 detected by ASKAP, localized to a galaxy at redshift 0.47.
%and found to pass through the halo of a foreground galaxy at redshift 0.37.
At microsecond time resolution the burst is resolved into four narrow pulses with a rise time of just $15\,\mu$s for the brightest.
The pulses have a diversity of morphology, but do not show evidence for temporal broadening by turbulent plasma along the line of sight, nor is there any evidence for periodicity in their arrival times.
%The pulses show lack of periodicity in their arrival times and have a diversity of morphology but do not show evidence for temporal scatter broadening by turbulent plasma along the line of sight.
The pulses are highly polarized (up to 95\%), with the polarization position angle varying both between and within pulses.
The pulses have apparent rotation measures that vary by $15\pm2\,\mrmunits$ and apparent dispersion measures that vary by $0.041\pm0.004\,\mdmunits$. Conversion between linear and circular polarization is observed across the brightest pulse.
We conclude that the FRB\,181112 pulses are most consistent with being a direct manifestation of the emission process or the result of propagation through a relativistic plasma close to the source.
This demonstrates that our method, which facilitates high-time-resolution polarimetric observations of FRBs, can be used to study not only burst emission processes, but also a diversity of propagation effects present on the gigaparsec paths they traverse.
%High time resolution polarimetric observations of future bursts can be used to not only study burst emission processes, but also a diversity of propagation effects present on the gigaparsec paths they traverse.
\end{abstract}

%% Keywords should appear after the \end{abstract} command. 
%% See the online documentation for the full list of available subject
%% keywords and the rules for their use.
%Radio bursts (1339), Astronomical techniques (1684), Radio interferometry (1346), Radio astronomy (1338)}

%% From the front matter, we move on to the body of the paper.
%% Sections are demarcated by \section and \subsection, respectively.
%% Observe the use of the LaTeX \label
%% command after the \subsection to give a symbolic KEY to the
%% subsection for cross-referencing in a \ref command.
%% You can use LaTeX's \ref and \label commands to keep track of
%% cross-references to sections, equations, tables, and figures.
%% That way, if you change the order of any elements, LaTeX will
%% automatically renumber them.
%%
%% We recommend that authors also use the natbib \citep
%% and \citet commands to identify citations.  The citations are
%% tied to the reference list via symbolic KEYs. The KEY corresponds
%% to the KEY in the \bibitem in the reference list below. 

\section{Introduction} \label{sec:intro}
The cause of the highly luminous, millisecond-timescale emission of fast radio bursts (FRBs) is not understood.  An increasing accumulation of bursts localized to host galaxies at cosmological distances, with redshifts over the range $0.19 < z < 0.66$ \citep{Chatterjee2017,Bannister2019,Ravi2019_localisation,Prochaska+19}, shows that their $\sim 10^{35}\,$K emission brightness temperatures are comparable to the coherent radiation of radio pulsars, but their $10^{29}$--$10^{33}$\,erg\,Hz$^{-1}$ spectral energy densities exceed those typically observed in pulsars by over ten orders of magnitude \citep{Shannon2018}.   

Efforts to identify the progenitors of FRBs and to explain the radio emission mechanism \citep{FRBtheoryCat} have been thwarted by the large set of uncertainties surrounding their basic physical properties.  The burst energetics are uncertain: what is the smallest timescale on which FRB emission occurs, and to what scale does this limit the size of the emission region and hence the volumetric energy density?  Analysis of FRB\,170827 \citep{Farah2018} shows that burst emission can exhibit complex temporal structure at the few-microsecond level, and that emission regions may be limited to sizes of only kilometers.  It is unclear which mechanism dominates the burstiness of the emission: it may be that the pulse duration is regulated principally by the emission mechanism itself, or it may instead be that the pulse envelope is governed by the motion of highly beamed emission across the line of sight.  High time resolution polarimetric observations could prove a decisive diagnostic of these systems in the same manner that similar observations revealed the systematic rotation of magnetic field lines through the line of sight in many pulsars \citep{RadhakrishnanCooke1969}.  Measurement of the periodicity in burst emissions, if this is a well-defined quantity, would further provide a crucial test of some emission models \citep[e.g.][]{Lyutikov19b}.  Finally, high time resolution studies of FRB emission may address whether or not there are analogues to FRB emission in the local Universe; for instance, \cite{Lyutikov19a} suggests that the time-frequency structure of some repeating FRBs resembles that of Type III Solar radio bursts.

%Despite the discovery of nearly 100 FRBs (see the FRB catalogue\footnote{\href{http://frbcat.org/}{http://frbcat.org/}}; 

%The voltage-capture system needed for interferometric localization of FRBs \citep[see][]{Bannister2019} on the Australian SKA Pathfinder (ASKAP) affords a means of addressing these questions in detail. This system consists of a ring buffer that stores the 3.1\,s of raw telescope data, which may be downloaded upon real-time detection of an FRB in the signal path for subsequent analysis.
%This system provides the capacity to reconstruct the burst signal in full polarization at the highest time resolution possible for the 336\,MHz bandwidth of the telescope beamformers (3 nanoseconds).  The system also typically boosts the S/N of the initial detection by at least an order of magnitude: initial ASKAP FRB detections are made from a data stream in which the total power signals of all antennas in the 36-element array are summed, but the coherent combination of data from each of the telescope's $N=12$--$36$ antennas boosts the S/N by a factor $\sim N^{1/2}$. A further S/N increase is realised if the burst width is considerably shorter than the $0.8-1.6\,$ms time resolution of the incoherent detection system.

In this paper, we present a high time resolution analysis of FRB\,181112 discovered as part of the Commensal Real-time ASKAP Fast Transients (CRAFT) survey \citep{Macquart2010}.
This burst was initially detected with a signal-to-noise ratio (S/N)  of 19.3 in the CRAFT incoherent detection pipeline, in observations centred at 1297.5\,MHz, with a measured duration of 2.1(2)\,ms and a fluence $26(3)\,$Jy\,ms, as reported by \cite{Prochaska+19}. 
The burst was localized to a host galaxy at $z=0.4755$ but, significantly, the line of sight also intercepted the halo of an intervening galaxy at $z=0.3674$ at a transverse distance of 28\,kpc.

In \S\,\ref{sec:method}, we provide an overview of the method used to produce the high signal-to-noise, high time resolution voltage time series. 
The temporal, spectral, and polarimetric properties of the burst are presented in \S\,\ref{sec:basichtr} and analyzed in  \S\,\ref{sec:analysis}. In \S\,\ref{sec:discussion}, we discuss the implications of these results on FRB emission theories and the properties of the media through which the burst propagated.

%% The "ht!" tells LaTeX to put the figure "here" first, at the "top" next
%% and to override the normal way of calculating a float position
\begin{figure*}[ht!]
\plotone{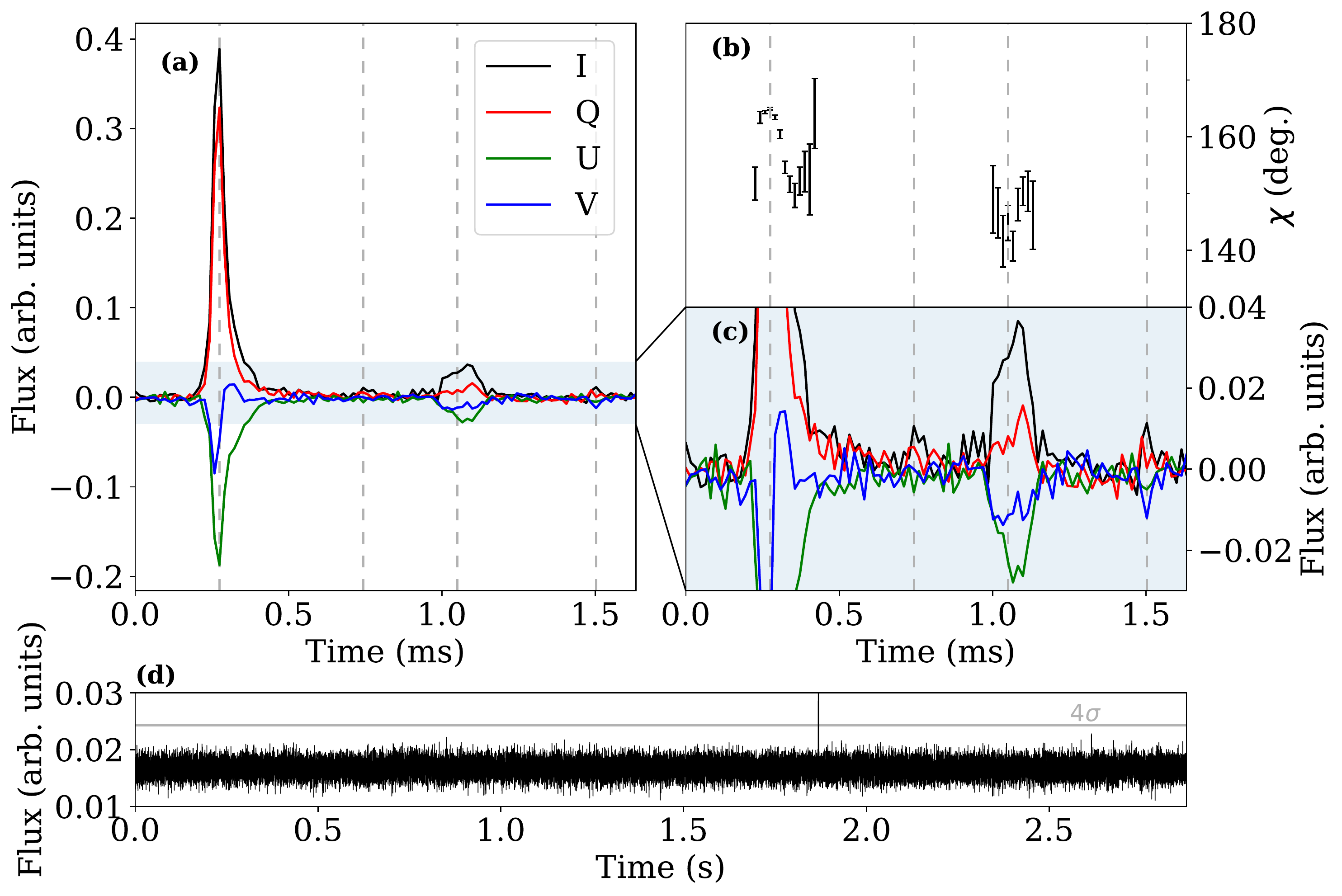}
\caption{Frequency-averaged burst profiles. The signal has been coherently dedispersed at ${\rm DM}= \dmval\,\mdmunits$ and corrected for Faraday rotation with ${\rm RM}=10\,\mrmunits$. The Faraday rotation was corrected with the linear polarization position angle referenced to the central frequency (\fc\,MHz). A total of 4 peaks are found at high time resolution, as indicated by the gray dashed vertical lines. (\emph{a}) Stokes parameters for the four pulses at $16\,\mu$s resolution; (\emph{b}) Relative polarization position angle $\chi$; (\emph{c}) Stokes parameters zoomed in near the baseline (shaded in blue) for a better view of the weaker sub-pulses; (\emph{d}) Intensity time series for the coherently dedispersed $\sim$3 second voltage buffer, where boundary regions have been discarded. 
%The plot excludes $\sim0.23$ seconds at the beginning and $\sim0.04$ seconds at the end of the buffer. 
The gray horizontal line shows a signal-to-noise ratio of 4.
\label{fig:all_in_one_time}}
\end{figure*}

\section{Voltage data processing methods} \label{sec:method}

The data presented here are derived from the ASKAP voltage capture system, which encodes the electric field saved by each antenna.
A detailed overview of the ASKAP voltage capture system is given in \citet{Bannister2019} and  \citet{Clarke2014}, and a brief summary of the relevant aspects of the high time resolution data reconstruction is presented here. The scripts implementing the following data processing steps are available in the CRAFT git repository\footnote{\url{https://bitbucket.csiro.au/scm/craf/craft.git}}. 

\subsection{Reconstruction to high time resolution via inverse polyphase filterbank}

The channelized voltage buffers are produced after a forward polyphase filterbank (PFB) \citep{Bellanger1976}. The ASKAP forward PFB is designed such that each of the $336$ coarse channels is $\sim$1\,MHz wide, which results in $\sim$\,$1\,\upmu$s time resolution. In order to increase the time resolution, a PFB inversion can be performed to synthesize a wider band and higher time resolution, which for the case of ASKAP is $(336~\rm{MHz})^{-1} \approx 3~\rm{ns}$.  
%This inverts the process of channelizing and hence the time resolution increases to the inverse of the total bandwidth. 
%In the case of ASKAP, since the total bandwidth of the triggered complex sampled voltage data is \bw\,MHz, the highest time resolution achievable is $(336~\rm{MHz})^{-1} \approx 2.976~\rm{ns}$.

Typically, PFB inversion is employed using a synthesis filterbank method \citep{Princen1986}.
However, a synthesis filterbank unavoidably leads to artifacts in the recovered signal unless the corresponding analysis filterbank (the forward PFB used here) uses a filter design that satisfies several restrictive criteria \citep{Vetterli1995:WSC}.  Since these criteria conflict with other desirable filter attributes, they were not satisfied in the case of the ASKAP PFB.
%However, unless the analysis filterbank (the forward PFB used here) is explicitly designed with inversion in mind, the synthesis filterbank approach unavoidably leads to artifacts in the recovered signal.
For an oversampled PFB such as that used by ASKAP, an alternate inversion method via Fourier transform can be used \citep{Morrison2019}. The Fourier transform inversion technique essentially discards the overlapping transition regions and applies an amplitude equalisation (``de-rippling") to correct for the analysis filterbank frequency response in the passband prior to a simple inverse Fourier transform.% In this paper, we apply the Fourier transform inversion method to each polarisation for each antenna separately, prior to the subsequent steps described below.

\subsection{Coherent beamforming and dedipsersion}\label{sec:method_beamform_dedisperse}

Taking advantage of access to the voltage data and the known sky position of the burst, we coherently sum (beamform) the received FRB signal. Prior to summation, the geometrical arrival time delays between the different dishes are removed and per-antenna amplitude, phase, and delay calibration terms derived using the calibration pipeline described in \citet{Prochaska+19} are applied to each antenna's Nyquist voltages.  Coherent beamforming is then simply the summation of the calibrated voltages.  
In addition, we use the coherent dedispersion technique \citep{Hankins1971,Hankins1975,Lormier2012book} to exactly correct for the dispersion of the signal. The dispersion measure (DM) used for dedispersion is determined with the PSRCHIVE tool {\tt\string PDMP}\footnote{\url{http://psrchive.sourceforge.net/manuals/pdmp/}} which identifies the DM that gives a maximal S/N for each pulse. 
%Considering FRB\,181112's narrow and bright pulse profile, optimizing the S/N also effectively maximizes the structure of the burst. 
Unlike FRBs that have been seen to repeat (e.g., \citet{Hessels2019}), we see no evidence of short-timescale structure in the profile of FRB\,181112 which becomes pronounced at a DM other than that which results in the maximum S/N.
The DM search is done with steps of $0.001\,\mdmunits$.

A coherently beamformed, dispersed dynamic spectrum of FRB\,181112 can be found in Fig.\,1(a) of \citet{Prochaska+19}.
Coherent beamforming significantly enhances the FRB S/N relative to that of incoherent beamforming (where the intensities of the individual stations are summed), while coherent dedispersion allows intrinsically shorter emission features to be recovered, further increasing S/N for initially unresolved pulses.
In the case of FRB\,181112, the S/N increases more than
a factor of ten when the voltages are both coherently
summed and coherently dedispersed from a S/N of $\snics$ to 
$\sntab$.

\subsection{Treatment of the corrupted data}

Of the 12 antennas used in the observation, one antenna (ak01) had missing data in eight 1~MHz highest frequency channels. For analysis of the spectral properties of the burst, the 8\,MHz bands were flagged after data from all 12 antennas were beamformed. For full time resolution analysis which needs the full bandwidth data, the problematic antenna (ak01) was not included.

\subsection{Polarization calibration} \label{sec:polcal}

%Prior to any polarimetric analysis, data from both polarizations were calibrated. This was done by observing the Vela pulsar with ASKAP, the polarimetric properties of which are well known. 
The data are polarization calibrated using an observation taken of the Vela pulsar (PSR~J0835$-$4510) 4 hours after the FRB was discovered.  As the properties of the Vela pulsar are well known, the observation could be used to determine instrumental leakage parameters (differential gain and phase between the two linearly polarized receptors), which could then be applied to the burst data set. Additional details of the polarization calibration method are described in \S S1.3 in \citet{Prochaska+19}. We note a difference in the Stokes parameters presented here and that presented in  \citet{Prochaska+19}. 
The change accounts for the handedness of the linearly polarized receptors of the phased array feed, which changes the signs of the Stokes parameters.  We note that the sign of Stokes-V follows IEEE convention, which is the standard for FRB and pulsar observations \cite[][]{vanStraten2010}. 

% Limit due the fact vela is detected in the different beam from the FRB
Uncertainty in the polarization fidelity is introduced by the relative position of the FRB in the ASKAP PAF beam pattern. The Vela observation was taken at beam center, while the FRB position, as measured interferometrically, was $20$~arcmin from beam center.  This is still well inside the half power point and ongoing analysis of ASKAP's instrumental polarization indicate that any additional calibration error is less than 2 percent.

\section{Properties of FRB\,181112} \label{sec:basichtr}

\begin{deluxetable}{lllll}
\tablenum{1}
\tablecaption{Pulse properties
\label{tab:summary_observation}}
\tablewidth{0pt}
\tablehead{
\colhead{Pulse} & \colhead{S/N}& \colhead{DM} & \colhead{RM} & \colhead{Arrival (ms)$^{\rm b}$}
}
%\decimalcolnumbers
\startdata
  1 & 220 &\dmval(1) &         \rmval(\rmerr) & 0 (reference) \\
  2 &  5 &\nodata &       \nodata  & 0.48(1) \\%0.624(6)\\ 
  3 & 28 &\trddmval(1) &    \trdrmval(\trdrmerr) & 0.808(4) \\ %0.787(2) \\
  4 & 8$^{\rm a}$ & \fthdmval(4)  &   \nodata & 1.212(2) \\ %1.387(3)\\
\hline
\enddata
\tablecomments{Values in parentheses represent $1\sigma$ uncertainties on the last digit. DM and RM uncertainties are each obtained from {\tt\string PDMP} and {\tt\string RMFIT}.\\$^{\rm a}$ The fourth pulse's S/N is measured after being dedispersed to its best fit DM (see \S~\ref{sec:dm}).\\$^{\rm b}$  See \S~\ref{sec:basichtr} for the explanation on arrival time measurement method.}
\end{deluxetable}

% dynamic spectrum of Stokes
\begin{figure*}[ht!]
\plotone{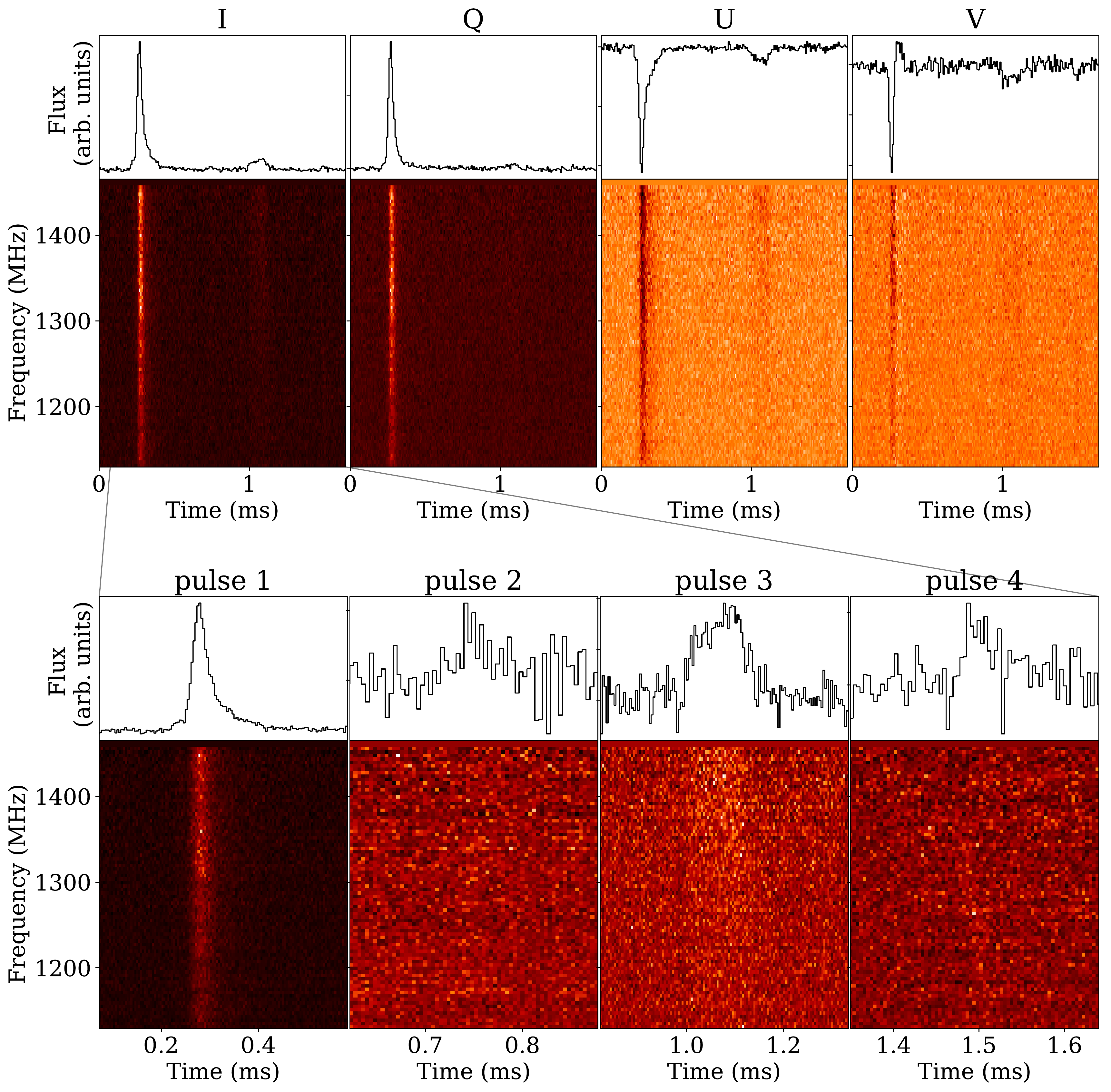}
\caption{Dynamic spectra of FRB\,181112 dedispersed at $\rm{DM_1}=\dmval\,\mdmunits$. (\emph{Upper panels:})  Spectra for all four Stokes parameters  at $8\,\mu$s temporal resolution and $4$\,MHz spectral resolution. (\emph{Lower panels:}) Cut-outs of total intensity (I) dynamic spectra around each pulse at $4\,\mu$s and $4$\,MHz resolutions.
\label{fig:all_dynamic_spec}}
\end{figure*}

Multiple components were identified in the frequency averaged burst time series, and the properties of each pulse are described in Table~\ref{tab:summary_observation}. The first and brightest pulse has a rise time of only $15\,\mu$s. The four pulses respectively have S/N of 220, 5, 28, and 8. Outside of the four pulses presented here, nothing exceeds the S/N threshold of 4.

%The pulses are visible at $t \approx 0.3$~ms, $0.75$~ms, $1.1$~ms, and $1.5$~ms as labeled in Figure~\ref{fig:all_in_one_time}(a)'s x-axis.
No periodic relationship was found in the pulse arrival times (shown in the fifth column of Table~\ref{tab:summary_observation}). To measure the pulse arrival times, we perform nested sampling using {\tt\string Bilby} (the same method described in \S\ref{sec:profile_fit}) instead of the standard pulsar arrival time measurement technique using the PSRCHIVE task {\tt\string paas}. The latter can fail for weaker pulses (such as pulses 2 and 4), where the pulse template may be a poor match to the actual pulse morphology. When fitting with {\tt\string Bilby}, each pulse is fitted with a Gaussian convolved with an exponential, and the arrival time is taken as the best-fit Gaussian's center. We note, however, that the uncertainty in pulse shape leads to an unavoidable additional uncertainty in defining pulse arrival times.
%To measure the pulse arrival times, we form analytic models of the individual pulses using the PSRCHIVE task {\tt\string paas}. The models are aligned to have zero phase between them. Time tagging is done with a Fourier domain monte carlo method, which provides more reliable uncertainties when the pulses are faint (as is the case for pulse 2 and 4). Nevertheless, we note that the uncertainty still exists in determining pulse arrival times associated with using a pulse template on the pulses with diverse morphology. 
No periodicity is found in the dedispersed time series using FFT-based searching with the PRESTO \citep{Ransom2001_PRESTO} routine {\tt\string accelsearch}. For the periodicity search, all candidates with S/N $> 3\,\sigma$ were folded and inspected by eye.

% polarization variation between pulses and across the first pulse
In addition to showing varying widths and intensities, 
the four pulses show different polarization properties, also varying across individual pulses.
 The Stokes parameters and polarization position angle (P.A.) in the time domain are shown in Figure~\ref{fig:all_in_one_time}.
 
The pulses also show different spectral structures.
The dynamic spectra of the pulses are shown in Figure~\ref{fig:all_dynamic_spec}. 
When dedispersed at the DM of the first component, the fourth pulse shows a residual time-frequency drift, suggesting that it either has a different DM than the other pulses, or that the burst emission drifts with frequency in a manner inconsistent with dispersion, as has been seen in repeating FRBs such as 121102 \citep{Hessels2019}. The low S/N precludes a definitive discrimination between these two possibilities.
%or exhibits nondispersive drift like repeat pulses from FRB\,121102 \citep{Hessels2019}.

% plots from FRB181112 Science paper
%\begin{figure}[ht!]
%\plottwo{FRB181112_prof.eps}{pa_comp.eps}
%\caption{plot from Science FRB181112 paper}
%\end{figure}

The dynamic spectra of the Stokes components are displayed 
in the top panels of Figure~\ref{fig:all_dynamic_spec}.
The burst shows evidence for signficant circular polarization, with the first pulse having a narrow circularly polarized component that switches sign and the fourth having large fractional circular polarization.

\section{Analysis} \label{sec:analysis}

\subsection{Dispersion measure} \label{sec:dm}

There is evidence that the dispersion of the fourth pulses is larger than the first and third. 
We fit for the fourth pulse's DM with the same method outlined in \S\ref{sec:method_beamform_dedisperse}. The best-fit DM value of the fourth pulse is measured to be $\rm{DM_4}= \fthdmval \pm \fthdmerr\,\mdmunits$. Thus, the DM difference between the first (and the brightest) pulse and the fourth pulse is
$$\Delta \rm{DM} = |\rm{DM_{1}} - \rm{DM_{4}}| = \pgfmathparse{\fthdmval-\dmval}\pgfmathresult\pm0.004\,\mdmunits.$$
The S/N increases from 6.0 when dedispersed with $\rm{DM_1}$ to 7.6 with its best-fit $\rm{DM_4}$ with a narrower pulse shape when only the channels containing strong signal ($\lesssim 1400\,{\rm MHz}$) are frequency averaged. The uncertainties are $1\sigma$.
%The two dynamic spectra for the fourth pulse in the lower panels of Figure~\ref{fig:all_dynamic_spec} confirm that the fourth peak is dedispersed better (signal aligned in time) with $\rm{DM_4}$.
%\begin{figure}[ht!]
%\gridline{\fig{c4_b4f4.pdf}{0.4\textwidth}{}
%            }
%\caption{Dynamic spectrum of the fourth pulses' intensity coherently dedispersed at two different dispersion measures (DM). (\emph{left}) Coherently dedispersed at $\rm{DM}=\dmval~\mdmunits$; (\emph{right}) coherently dedispersed at $\rm{DM}=\fthdmval~\mdmunits$.  Both plots are shown in $4\,\mu$s temporal resolution and $4$\,MHz spectral resolution.
%\label{fig:p4dm}}
%\end{figure}

In addition, the DM of the third pulse is measured to be $\rm{DM_{3}}=\trddmval\pm\trddmerr\,\mdmunits$, which is consistent with the first pulse's DM. The larger uncertainty is a result of its broad pulse shape.

\subsection{Rotation measure} \label{sec:rm}

There is evidence for differences in rotation measure between pulses 1 and 3, for which we detected significant linear polarization.
We estimated the rotation measures of the pulses using the \texttt{RMFIT}\footnote{\url{http://psrchive.sourceforge.net/manuals/rmfit/}} program, which is part of the \texttt{PSRCHIVE} software package \cite[][]{Hotan2004}. The program determines the rotation measure by maximizing the frequency-averaged fractional linear polarization.
The two measured RM values disagree with each other. The  RM of the first pulse is $\rm{RM}_{1} = \rndrmval \pm \rndrmerr\,\mrmunits$, while the RM of the third pulse is $\rm{RM}_{3} = \rndtrdrmval \pm \rndtrdrmerr\,\mrmunits$.

\begin{figure}[h!]
%\gridline{\fig{PAspec_p1.pdf}{0.4\textwidth}{(a) 1st pulse, RM=$10\,\mrmunits$}
%          }%0.45
%\gridline{\fig{PAspec_p3.pdf}{0.4\textwidth}{(b) 3rd pulse, RM=$26\,\mrmunits$}
 %         }%0.45
\plotone{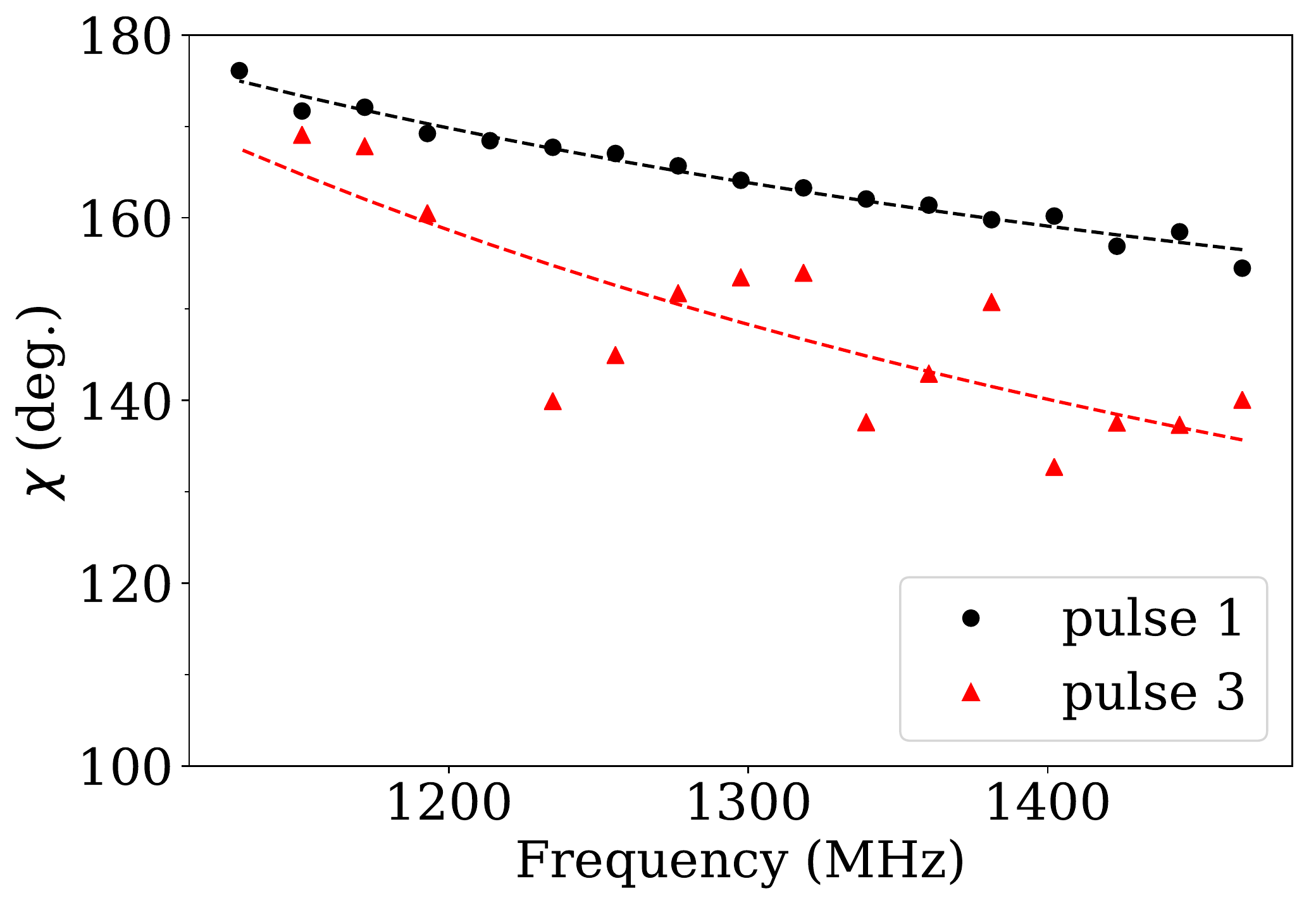}
\caption{Polarization position angle $\chi$ as a function of frequency.
The black and red dots indicate the measured $\chi$ for pulse 1 and pulse 3, respectively, and the dotted lines indicate the fitted curve $\chi\propto\lambda^2$. The data points shown here have S/N of above $1.5\,\sigma$.
%The black data points indicate $\chi$ prior to Faraday rotation correction and red solid lines indicate $\chi$ after Faraday rotation correction with each the best fitting rotation measures (RM) for each pulse. (\emph{a}) The $\chi$ of the first pulse before and after correcting for Faraday rotation with $\text{RM}=10\,\mrmunits$; (\emph{b}) the third pulse's $\chi$ before and after correcting for Faraday rotation with $\text{RM}=26\,\mrmunits$.
\label{fig:rmdiff}}
\end{figure}

The significant difference in rotation measure is evident when the linear polarization position angle, $\chi$, is measured across the band. Figure~\ref{fig:rmdiff} shows the uncorrected positional angles for the two pulses.

It is noteworthy that \citet{Lu2019} predicts smaller DM and RM  for the brighter pulse due to strong-wave effects in plasma in the vicinity of the burst source, which is partially consistent with our observations.  The faint fourth pulse has a larger apparent dispersion than the first pulse.   While the fainter third pulse shows a larger rotation measure than the first pulse, it has consistent dispersion measure with the first pulse.
%  Figure~\ref{fig:rmdiff}(a) shows the position angle of the first pulse, corrected to  a reference  $\rm{RM} = 10\,\mrmunits$ which is close to the best-fit value of $\rm{RM}_{1} = \rndrmval\,\mrmunits$.
% After the correction, $\chi$ is constant with frequency.
% Similarly when the third pulse is corrected to $\rm{RM} = 26\,\mrmunits \approx \rm{RM}_{3}$, $\chi$ is constant with frequency.
% However, is the observation ever consistent with $\chi \propto \lambda^2$? This is a point to which we return in the discussion \S~\ref{sec:rel_plasma}.

\subsection{Polarization properties}

 % Poincare sphere
\begin{figure}[ht!]
\gridline{\fig{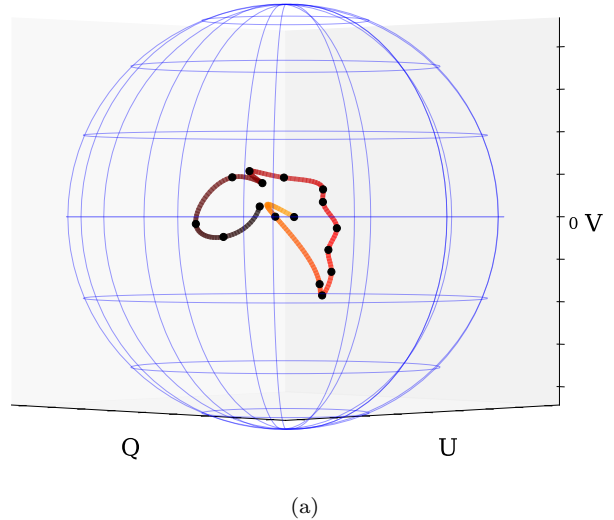}{0.465\textwidth}{(a)}
            }
\gridline{\fig{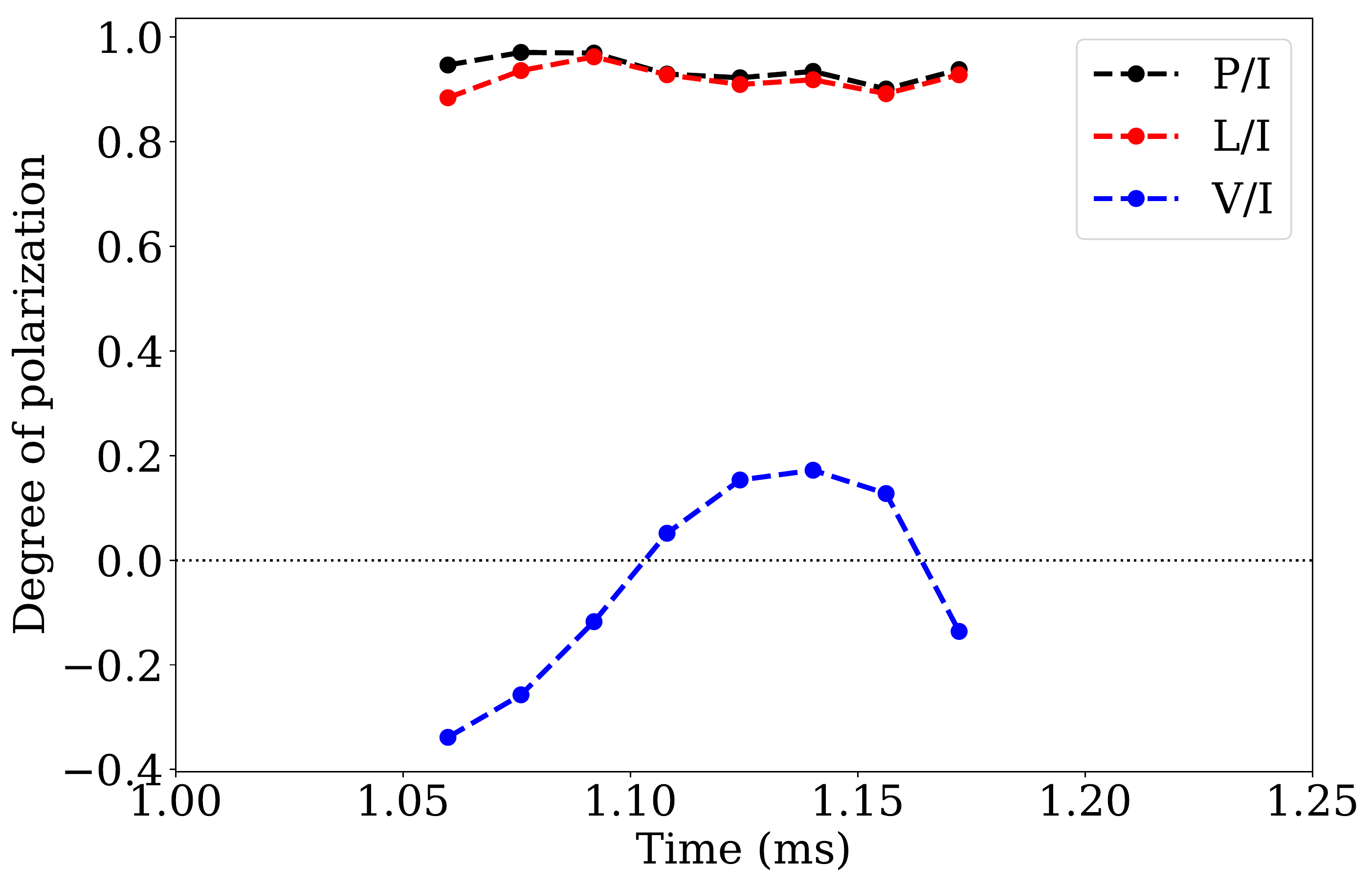}{0.465\textwidth}{(b)}
            }         
%\plotone{poincare_t.pdf}
\caption{
The first pulse's polarization state in two different representations. (\emph{a}) Polarization mapped on to the \Poincare sphere. The black dots are the data points and the colored line is an interpolation between points. Evolution of time within the pulse is represented in color starting from orange and ending in black in $10\,\mu$s resolution. (\emph{b}) Degree of polarization across the pulse at $16\,\mu$s resolution. Only data points above S/N $> 13\sigma$ are shown.
\label{fig:poincare}}
\end{figure}

The first pulse has a high polarization fraction of $P/I = \sqrt{L^2+V^2}/I\sim 0.94$, which remains approximately constant across the pulse. Figure~\ref{fig:poincare} shows that the polarization states remain closely near the surface of the \Poincare sphere as time evolves.
It is very significant that while the total polarization fraction stays constant, the circular and linear polarization fraction varies substantially across the pulse. This, along with the polarization fraction forming a closed loop, leads to speculation of relativistic plasma propagation, presented in \S~\ref{sec:rel_plasma}.
The average linear polarization fraction across the pulse is  $\langle L/I\rangle_t \sim 0.92$, with the maximum value being $L/I = 0.96$. The degree of circular polarization $V/I$ shows a significant variation across the main pulse, ranging from $-0.34 < V/I < 0.17$.
The debiasing step of $P$ and $L$ is omitted because the bias is negligible due to the high S/N of the pulse.

% HC: Maybe we can move this subsection after the polarization properties and before the scintillation, so that the two methods (this profile fitting and ACF) of giving scattering constraints are right after another.
\subsection{Burst morphology}\label{sec:profile_fit}
We use Bayesian methodology to model the burst shape and spectral variations of the first and third pulses in detail.
Because the second and fourth pulses are weak, we do not include them in this analysis.
For the total intensity time series for each of the two modeled pulses, we fit models consisting of one or two Gaussian components, optionally convolved with an exponentially decaying scattering tail. Deviations from the overall best-fit dispersion measure are also fit.

To undertake this analysis, we subdivide both pulses into $8$ subbands, which are then modelled using the nested sampling method \texttt{Dynesty} \citep{dynesty} implemented in \texttt{Bilby} \citep{bilby}.

\begin{deluxetable*}{lcccccc}
\tablenum{2}
\tablecaption{Model comparison for the first pulse.
\label{tab:pulse profile}}
\tablewidth{0pt}
\tablehead{
\colhead{Model} & \colhead{DM offset} &\colhead{$\sigma_1\,^{\rm a}$}&\colhead{$\sigma_2$} & \colhead{$\alpha$}& \colhead{$\tau^{\rm b}$} & \colhead{RMS error$^{\rm c}$}%\colhead{Bayes factor$^{\rm b}$}
\\
&($\times 10^{-3}$ pc cm${^{-3}}$)&($\mu$s)&($\mu$s)&
&($\mu$s)&}
%\decimalcolnumbers
\startdata
Scattering (SGE, $\alpha=-4$)& 6.8 $\pm 0.6$ & 15.8 $\pm 0.5$&--&--& $20.7^{+0.9}_{-0.8}$&2.7 \\%0.30\\%(reference)\\%$-1117.75$\\% \pm 0.38$\\
Scattering (SGE, $\alpha$ unconstrained) & 1.6 $\pm 0.5$& 13.6 $\pm 0.3$&-- & $-2.0 \pm 0.3$& 24.6 $\pm0.6$ &2.1 \\%0.23\\%137.49\\
%$-1031.45$\\% $\pm 0.38$\\
Double Gaussians (DG)& $0.3 \pm0.4$ &$17.9\pm0.5$&$43.7^{+1.5}_{-1.5}$&--&--&2.1\\%0.23\\%144.06\\%$-966.35$  %$\pm 0.45$
\enddata
\tablecomments{$^{\rm a}$ $\sigma_i$ represents the width of the $i$-th Gaussian used in each model.
\\$^{\rm b}$ The frequency-dependent broadening timescale $\tau$ presented here is the value at the central frequency.\\
$^{\rm c}$ The RMS error is normalized with respect to the off-pulse RMS.}
%For reference, the off-pulse RMS is 0.11 in all models.}
%The relative Bayes factor is in logarithmic scale, representing how each model is preferred over the first model.}
\end{deluxetable*}
% Feb. 4) -180.658, -19.326, -38.381

The first component comprises a bright peak with fading postcursor emission. The morphology of the tail is qualitatively similar to the exponential tail commonly associated with multipath propagation but detailed analysis shows that the shape is not consistent with a pure exponential.  
We compare two models to characterize the first peak. 
In the first model, we assume the pulse tail is the result of scattering, so the pulse is modeled to be a Gaussian convolved with an exponential function (hereafter referred to as SGE, for single Gaussian plus exponential).
The broadening time $\tau$ is allowed to vary with frequency,  $\tau \propto \nu^{\alpha}$. The spectral index, $\alpha$, is both modelled as a free parameter or fixed with $\alpha \approx -4$ expected for multipath propagation in cold plasma.
For the second model we assume the pulse can be modelled with two Gaussians (DG, for double Gaussian), to account for the extended tail following first pulse.
The results are listed in Table \ref{tab:pulse profile}. 
We compare the models using their normalized root-mean-sqaure (RMS) error (i.e. RMS error in the region of the pulse divided by the off-pulse RMS) in the last column of the table. A higher value of normalised RMS error indicates that the model is relatively disfavored.
%relative logarithmic Bayesian evidence. The logarithmic Bayes factor in the last column of the table is simply the difference between the evidence. A positive Bayes factor indicates the model is favored than the reference model.
%The fit residual is displayed in Figures \ref{fig:scat_residual} and \ref{fig:dbb_residual}.
%Model 1 measures a scattering tail of 22 us at 1.272 GHz, the centre frequency of the data. 
%However, comparison of the two models' Bayesian evidence gives preference to Model 2 with high confidence.

%The DG model is formally preferred with a significant logarithmic Bayes factor of $\sim 7$ relative to the SGE $\alpha$-free model and $\sim 144$ relative to the SGE cold plasma $\alpha = -4$ model although the difference may result from neither the Gaussian nor the exponential being the correct shape.
The SGE cold plasma $\alpha = -4$ model is excluded as the alternate models have a decisively lower
RMS error of 2.1. 
However, it is worth noting that both the DG and SGE $\alpha$-free models still have significantly higher RMS compared to the off-pulse noise level and have residuals that indicate neither are the correct shape.
%The small RMS error difference between the DG and SGE $\alpha$-free models may result from neither the Gaussian nor the exponential being the correct shape.
Also note that a model fit with two components is just a description of the pulse shape and does not imply the emission is the physical superposition of two independent components, as this would not fit the polarization observations discussed in the preceding section without a region of depolarization in between.
The SGE models the best-fit frequency dependence for the pulse broadening with the frequency index of $\alpha =-2.0 \pm0.3$, far from the value of $-4$.
Both of these results disfavor multipath propagation in a cold plasma as the origin of the postcursor tail in the first pulse, meaning that the best-fit value for $\tau$ of 21 $\mu$s from the SGE model likely represents an upper limit to the actual amount of scattering exhibited by the pulse.  %A final piece of supporting evidence to this interpretation comes from the extremely rapid rise time of this pulse (15 $\mu$s).
%Both models also account for pulse misalignment with a DM offset of $\sim1\times 10^{-3}$ pc cm$^{-3}$  

We perform a similar analysis for the much wider third peak, comparing a SGE model with a single Gaussian model with no exponential tail to search for evidence of scattering. The model including an exponential tail was not preferred over a single Gaussian model alone, and hence the third pulse shows no evidence for scatter-broadening. A DG model is not considered because the third pulse shows a flat profile that is qualitatively dissimilar to the model. 

%{\bf Why no DG model here?  Why no quantitative values?}

%\begin{figure*}[ht!]
%\gridline{\fig{achromaticSTOKES_I_peak1_residuals.png}{0.8\textwidth}{}
%            }
%\caption{Residual of scattering model fit of first pulse
%\label{fig:scat_residual}}
%\end{figure*}
%\begin{figure*}[ht!]
%\gridline{\fig{STOKES_I_peak1_residuals.png}{0.8\textwidth}{}
%            }
%\caption{Residual of double Gaussian model fit of first pulse.
%\label{fig:dbb_residual}}
%\end{figure*}

\subsection{Constraints on scintillation}\label{sec:scatt_analysis}

The frequency auto-covariance function \cite[ACF, ][]{Cordes1983ApJ_ACF,Cordes1986ApJ_ACF} provides a second method for constraining the scattering timescale.
%The constraint on the scattering timescale is inferred to be $\mdtpulse \lesssim \tscatt\,\mu$s from the scintillation bandwidth.
The scintillation bandwidth is measured by constructing the ACF
\begin{equation}
    \mathrm{ACF}(\delta \nu) = \frac{1}{N} \sum_{\nu} \Delta S(\nu) \Delta S(\nu + \delta \nu),
\end{equation}
where $S(\nu)$ is the spectrum and $\Delta S(\nu) = S(\nu)-\bar{S}$, with $\bar{S}$ being the mean spectral power and $N$ being the number of frequency bins. The ACF is then fitted with a Lorentzian function \citep{Chashei1976_ACFfit}
\begin{equation}
    f(\delta \nu) = C \left( 1+ \frac{\delta \nu^2}{\delta \nu_d^2} \right)^{-1},
\end{equation}
where the two parameters $C$ and $\delta \nu_d$ are, respectively, a proportionality constant and the decorrelation (scintillation) bandwidth.

% ACF and its fit
\begin{figure}[ht!]
\plotone{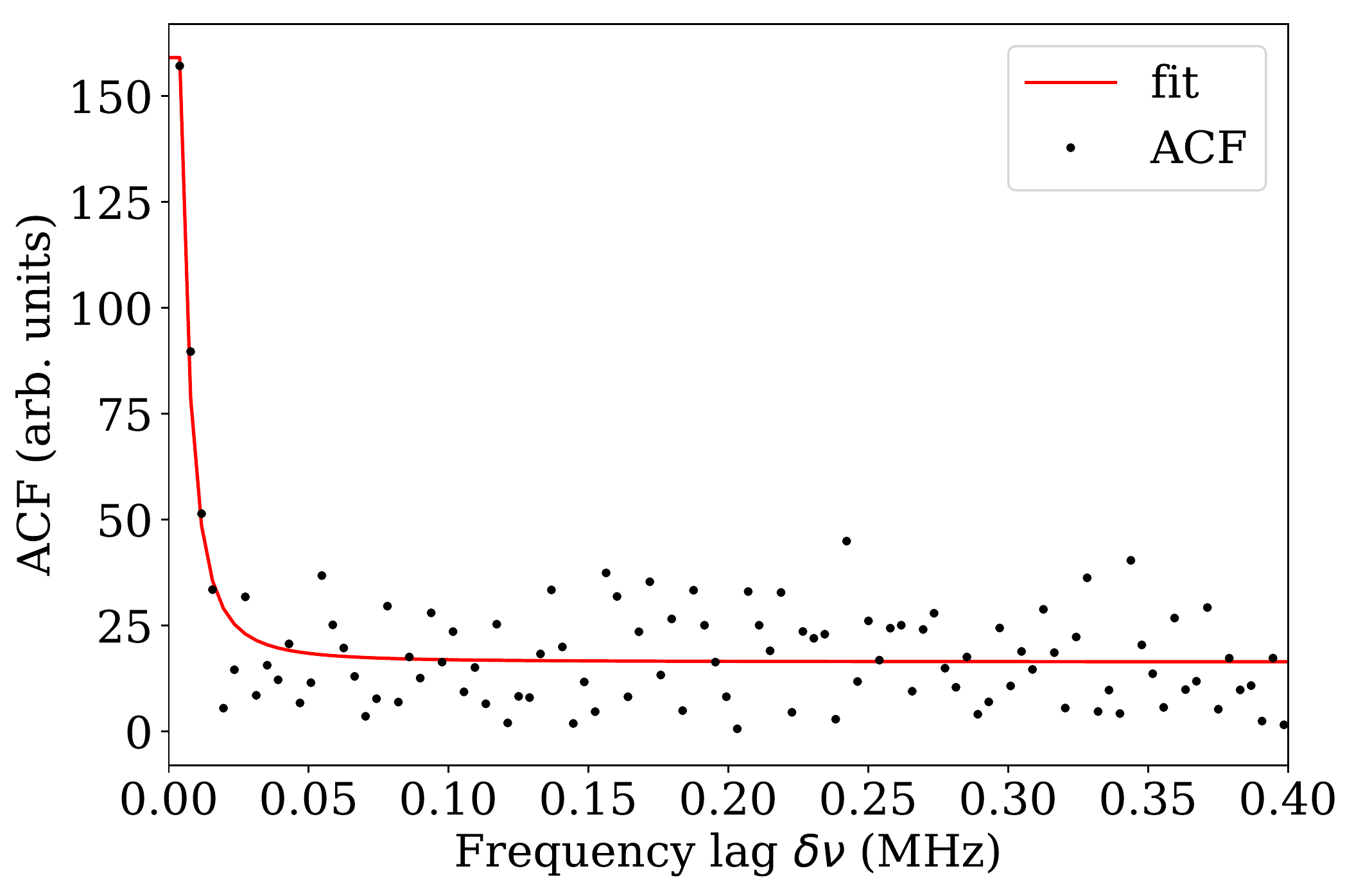}
\caption{
Measured ACF in the frequency domain and the fitted function for the main pulse of FRB\,181112. The fitted decorrelation bandwidth is $\delta \nu_d \sim \freqdc$ kHz. The zero-lag value of the ACF which is related to self-noise is not shown.
\label{fig:acf}}
\end{figure}

The main pulse's measured ACF and fitted function are shown in Figure~\ref{fig:acf}. We focus solely on the brightest pulse, as the S/N of the other three pulses is too low for a useful analysis. The decorrelation bandwidth of the main pulse is measured to be $\delta\nu_d =\freqdc\pm0.7 \,{\rm kHz}$. The decorrelation bandwidth $\delta\nu_d$ and the scatter broadening time $\mdtpulse$ follows a relation
\begin{equation}
    2\pi\,\delta\nu_d\,\mdtpulse = C_1,
\end{equation}
where $C_1 =0.654$ under the assumption that the scattering is due to Kolmogorov turbulence situated on a thin scattering screen \citep{Lambert1999}.  
This corresponds to a timescale of $\tau = 0.654\,(2\pi\,\delta\nu_d)^{-1} \sim \tscatt\pm4\,\mu$s. The timescale $\tau$ from the ACF analysis does not show strong frequency dependence, with different frequency subbands all having a value of $\sim 20\,\mu$s, so it is inconclusive that this time broadening is purely from propagation in a cold plasma. Thus, it can be interpreted that either (\textit{i}) the measured decorrelation bandwidth (approximately the resolution used for this analysis) is an upper limit $\delta\nu_d \lesssim \freqdc\,{\rm kHz}$ or (\textit{ii}) the scintillation bandwidth is too large to be detected ($\delta\nu_d > 100\,{\rm MHz}$). The corresponding constraint on the scattering timescale is then (\textit{i}) $\mdtpulse >\tscatt\,\mu{\rm s}$ or (\textit{ii}) $\mdtpulse < 0.001\,\mu$s.% < 40\,\mu$s. The upper bound of $40\,\mu$s is from \citet{Prochaska+19}, which is determined via direct measurement.
%the calculated $\tau$ can be used as a new upper limit on the scattering timescale $\mdtpulse \lesssim \tscatt\,\mu$s.

We can now revisit the scattering analysis of \citet{Prochaska+19}, where the temporal burst profile was modeled with relatively coarse time resolution ($54 \mu$s), placing an upper limit of $40\,\mu$s.  We repeat this approach but with much finer time resolution in \S~\ref{sec:profile_fit}, finding that the scattering time is $\sim 21\,\mu$s from the SGE $\alpha=-4$ fit.  We also have complementary information in the form of the frequency ACF which was not available to \citet{Prochaska+19}. The frequency ACF suggests that either the scattering time is $\gtrsim 25\,\mu$s, or else $\lesssim 1\,$ns.  These two complementary approaches are in moderate tension; either (\textit{a}) the scattering time is $\sim 20\,\mu$s and both the time domain fitting and frequency ACF fitting are slightly biased, or (\textit{b}) the scattering time is extremely small ($\lesssim 1\,$ns) and the time domain fitting is badly biased.  We consider the first alternative more likely, but in either case, the scattering time has been constrained to be at least a factor of $\sim 2$ lower than the upper limit presented by \citet{Prochaska+19}.
The limit of the density of the galaxy halo intercepted by the sightline of FRB\,181112, assuming a Kolmogorov spectrum of turbulence, is given as (\textit{a})
\begin{equation}
\begin{split}
\langle &n_e \rangle \sim 1.8 \times 10^{-3} \alpha^{-1} \\
&\times\left( \frac{\Delta L}{50\,{\rm kpc}} \right)^{-1/2} \left( \frac{L_0}{1\,{\rm kpc}} \right)^{1/3}  \left(\frac{\mdtpulse}{20 \, \rm \upmu s} \right)^{5/12}\,{\rm cm}^{-3},
\label{eqn:scatteringconstraint1}
\end{split}
\end{equation}
or (\textit{b})
\begin{equation}
    \begin{split}
\langle &n_e \rangle < 0.03 \times 10^{-3} \alpha^{-1} \\
&\times\left( \frac{\Delta L}{50\,{\rm kpc}} \right)^{-1/2} \left( \frac{L_0}{1\,{\rm kpc}} \right)^{1/3}  \left(\frac{\mdtpulse}{0.001 \, \rm \upmu s} \right)^{5/12}\,{\rm cm}^{-3} .\label{eqn:scatteringconstraint2}\\
    \end{split}
\end{equation}
The parameters in equations~\eqref{eqn:scatteringconstraint1} and \eqref{eqn:scatteringconstraint2} are defined in \citet{Prochaska+19}, with $\alpha$ in the above equations being different from the frequency exponent $\alpha$ defined in \S~\ref{sec:profile_fit}.

%We can combine the measured scattering timescale of $\tau\sim20\mu$s from Table~\ref{tab:pulse profile}, we are only left with case (\textit{i}).
%Comparing the case (\textit{i}) constraints with the equation (S21) of \citet{Prochaska+19}, stating that $\langle n_e \rangle < 2.3 \times 10^{-3} \alpha^{-1}  \left( \frac{\Delta L}{50\,{\rm kpc}} \right)^{-1/2} \left( \frac{L_0}{1\,{\rm kpc}} \right)^{1/3}  \left(\frac{\mdtpulse}{40 \, \rm \upmu s} \right)^{5/12}  
%\,{\rm cm}^{-3}$, we have a significanly tighter constraint on the intercepted galaxy halo. % 2.8

%Moreover, the timescale $\tau$ did not show strong frequency dependence, all having a value of $\sim 20\,\mu$s for different frequency subbands.
%This is again consistent with the results from an independent method in \S~\ref{sec:profile_fit}. Assembling this information, we can now rule out the multipath propagation in cold plasma scenario where the pulse broadening time shows a strong frequency dependence of about $\tau\propto\nu^{-4}$. Thus this leaves intrinsic mechanism of the burst, gravitational lensing scenarios, or relativisitic plasma lensing for explaining such pulse broadening.

\subsection{No correlation between two brightest pulses} \label{sec:corr}

If the presence of multiple pulses was produced by a propagation effect due to either gravitational or plasma lensing, we would expect spatial coherence of the emission, which would manifest as correlation in the voltages between pulses. 
%The voltage cross-correlation of the two brightest pulses, pulse one and pulse three, is computed to search for any correlation between the two pulses. If a significant cross-correlation is present, implying coherence between the electric field measured at the telescope at the times of these pulses, it would strongly support the gravitational/plasma lensing scenarios as explanations of the separated sub-pulses.
%Initially, we applied a simple cross-correlation to the coherently de-dispersed datastream, finding no significant correlation around a time separation corresponding to the separation between pulse 1 and pulse 3. 
When cross-correlating the voltages from the pulses, it is essential to search both over possible time lags and dispersion measure differences, as even very small amounts of differential dispersion can cause phase variations large enough to decorrelate the signals. We searched for correlation between pulse 1 and pulse 3, following the method described in detail in \citet{Farah2019}.  
The signal of the third pulse is dedispersed with a range of trial values prior to cross-correlation.  The trial DM values range from $\rm{DM_{1}}-0.01\,\mdmunits<\rm{DM}<\rm{DM_{1}}+0.01\,\mdmunits$ with DM steps of $10^{-6}\,\mdmunits$ and time lags between $-32\,\mu$s and $32\,\mu$s. 
No correlation is found. However, no observation of correlation does not completely rule out the multipath propagation scenarios since slightly different scattering screens in each path can easily destroy the spatial coherence.

\subsection{Microstructure}

The shortest time scale fluctuations (referred to as microstructure) in the pulse intensity place important constraints on the emission region size and hence the emission process. We observe no microstructure on timescales less than the scattering timescale of $\approx 20\,\mu$s using the two techniques described here.

First, intensity power spectra are computed at different time resolutions and compared to determine if power is detected at higher time resolutions. The intensity power spectrum, $P_{\rm high}$, at the highest time resolution is computed as a Fourier transform of intensity at the highest time resolution $P_{\rm high} = \mathcal{F}\{I_{3\,\rm ns}(t)\}$. Then the intensity power spectrum at a lower time resolution, $P_{\rm low}$, is added with the power spectrum of off-pulse noise computed at the highest time resolution $P_{\rm high, off}$. If $P_{\rm high}$ shows an extra power compared to $P_{\rm low}+P_{\rm high, off}$, this would imply that this FRB exhibits short-timescale intensity variations. However, we are not able to find a significant difference between the two spectra in the case of the brightest pulse. We computed $P_{\rm low}$ with resolutions ranging from $1\,\mu\rm s$ to $64\,\mu\rm s$, and also attempted applying the method after dividing the $\bw\,\rm MHz$ into 2, 4, and 8 subbands to search for extra variance in each subband, but nothing was found.

In the second method, we used the temporal autocovariance function \citep{Hankins1972,Lange1998},  $\mathrm{ACF}(\delta t) = C \sum_{t} \Delta I(t) \Delta I(t + \delta t)$, where $C$ is a normalization constant and $\Delta I$ is the intensity at each time where the mean of the off-pulse intensity is subtracted. This method is applied to the two brightest pulses, pulse 1 and pulse 3. We were not able to find significant evidence of microstructure in either pulse.

It is possible that temporal smearing due to multipath propagation, for which the timescale is estimated to be $\approx 20\,\mu$s (\S~\ref{sec:scatt_analysis}), has obscured
the shortest timescale burst fluctuations.

\section{Discussion} \label{sec:discussion}

Several scenarios can account for some of the observed properties of FRB\,181112.
In Table~\ref{tab:discussion_summary}, we summarize the set of models we consider and indicate if they can explain the burst properties discussed above.
%The full set of models and how each of them can or cannot explain the observed properties of the burst is summarized in %Out of those, the intrinsic emission mechanism and the gravitational lensing cases are discussed in depth in the following subsections.

% h for hidden column, D for decimal point arangement
\begin{deluxetable*}{ccccc}
\tablenum{3}
\tablecaption{Summary of how each scenario can or cannot explain the observed properties of FRB\,181112
\label{tab:discussion_summary}}
\tablewidth{0pt}
\tablehead{
\nocolhead{Common} & \colhead{Intrinsic emission mechanism} & \colhead{Cold plasma} & \colhead{Gravitational lensing} & \colhead{Relativistic plasma}
}
%\decimalcolnumbers
\startdata
Frequency structure & \checkmark & \checkmark & \nodata & \checkmark  \\
%Presence of circular polarization$^1$ & \checkmark & \nodata & \nodata & \checkmark \\
Polarization state conversion$^{\rm a}$  & \checkmark & \xmark & \xmark & \checkmark \\
%P.A. variations & \checkmark & \checkmark & \checkmark & \checkmark \\ 
P.A. swing &\checkmark  & \nodata & \nodata & \checkmark \\ 
DM variations$^{\rm b}$& \xmark & \checkmark & \checkmark & \checkmark \\ 
RM  variations & \checkmark &\checkmark & \checkmark & \nodata \\ 
No spatial correlation between pulses$^{\rm c}$ & \checkmark & \xmark & \xmark & \checkmark \\ 
%No spectral correlation between pulses & \checkmark &  \xmark & \xmark & \checkmark \\
%Frequency-independent exponential tail & \checkmark & \xmark & \checkmark &  \xmark \\ 
\enddata
\tablecomments{Please see the text for a more nuanced interpretation of each observational results. ``\nodata'' indicates that it is inconclusive whether a scenario can explain a corresponding property.\\$^{\rm a}$ Cold plasma and gravitational lensing cannot convert between linear and circular polarization.\\$^{\rm b}$ There could be an intrinsic change in time with frequency that would be misinterpreted as changing DM.\\$^{\rm c}$ The spatial correlation between pulses is only possible in the case of lensing but the absence of correlation does not rule out the lensing scenarios (see \S\ref{sec:corr}).}
\end{deluxetable*}

\subsection{Gravitational and plasma lensing}

Given the presence of multiple pulses
and the passage of the burst through the halo of the galaxy in the foreground, we first assess what properties of FRB\,181112 are consistent with being produced by propagation effects induced by non-relativistic plasma lensing or gravitational lensing.

A single pulse, incident on inhomogeneous plasma in the foreground galaxy halo could break up into multiple pulses (corresponding to multiple images), each with different arrival times \citep{Cordes2017}. 
The propagation paths could have different dispersion and rotation measures, which would explain these observed differences.
However, a non-relativistic plasma could not produce variations in circular polarization within a pulse. 
We would also expect the pulses to show spatial cross-correlation, which we do not observe.

% RMS: The two paragraphs need to be written in the same style

Similarly, a gravitational lens can produce multiple images of a background source, each with a different time of arrival, due to the effect of gravitational time dilation and the geometric path difference. Lensing by low mass (30-100 M$_\odot$) compact objects can result in time delays within our 3 second observing window. In the case that an FRB is lensed, we might expect disagreements in DM and RM (but not the intrinsic polarization) between the multiple components. %, however if the delay is $\lesssim 1\,$s these differences are likely unobservable.
The significant differences in the polarization properties of the two brightest pulses cannot be easily explained by this model.
As with plasma lensing, we would expect the pulses to show spatial correlation which we do not observe.
Further discussion will be provided in the upcoming paper (Sammons et al., in prep).

%Moreover we do not find fringes between the two peaks where their voltages are cross-correlated, which would be expected if the pulse structure were attributable to multipath propagation. (The fringes may be destroyed if the paths are separated sufficiently that they encounter different plasma scattering, so that the phases of each pulse are scrambled as a function of frequency.  However, the delay is so small that this is unlikely.)

%Furthermore multipath propagation through a non-relativistic plasma cannot produce the observed  difference  in circular polarization between the peaks.  It is unlikely therefore, that lensing has been observed in this case.  

\subsection{Intrinsic emission mechanism} \label{sec:emission_mech}

The short duration of the pulses in FRB\,181112 and the burst luminosity distance of 2.70\,Gpc imply a high energy density in the emission region. 
The fluence of the leading (strongest) pulse is measured to be 20.2(1)\,Jy\,ms, with a characteristic pulse timescale of $w=15\,\mu$s.  For a source size $\sim c w$ and attributing any beaming to Doppler boosting, the implied apparent energy spectral density is $9.0 \times 10^{13} $J\,m$^{-3}$\,Hz$^{-1}$ in the rest frame of the emitting source. The total energy density, under the conservative assumption that the emission is confined only to the 336\,MHz bandwidth over which it was detected, is $3.0 \times 10^{22}\,$J\,m$^{-3}$, equivalent to the energy density in a magnetic field $B = 3 \times 10^{12}\,$G.
% For the foregoing arguments, relativistic motion beams into a solid angle ~1/Gamma, but dV_observer = Gamma dV_rest frame, so (1/dVobs) = (1/Gamma) (1/dVrest).

The system also exhibits rapid polarization changes, on timescales comparable to the burst substructure. In the first pulse, we observe a change in the linear position angle of $\Delta\chi \approx 20\,$deg and the circular polarization variability on a timescale of only $40\,\mu$s. 
The third pulse shows significant differences in the mean circular polarization fraction and polarization profile from the first pulse (see Figure~\ref{fig:all_in_one_time}). However, the third pulse is insufficiently bright to determine whether it exhibits comparable position angle variations.

An obvious possible interpretation would link the position-angle variations to the magnetic field geometry in the burst emission region, either due to rapid evolution of the magnetic field topology over the burst duration or to rotation of the emission region and its associated magnetic field across the line of sight. Intrinsic changes would require substantial reconfiguration of the magnetic field on timescales of $\approx 40\,\mu$s.  However, if the emission region rotates across the line of sight (analogous to a radio pulsar) one need only invoke a static or slowly-varying field to explain the temporal polarization variability.  

The rotating vector model \citep[RVM; ][]{RadhakrishnanCooke1969} and its various extensions \citep[e.g.][]{Blaskiewicz91,HibschmanArons01,Lyutikov16} are commonly invoked to explain the regular linear polarization position angle swings observed in a substantial fraction of the pulsar population and could provide a framework for interpreting the polarization properties of FRB\,181112.  Neglecting aberrations, the maximum rate of position angle ($\chi$) swing with respect to pulse longitude ($\phi$) is $d \chi/d\phi  = \sin \alpha/ \sin \beta$, where $\alpha$ is the inclination of the magnetic field to the spin axis and $\beta$ is the impact angle relative to the line of sight.  The maximum rate of $\chi$\,\,change of $-189^\circ$\,ms$^{-1}$ observed in pulse one yields a spin-period, $P$, limit on the geometry of the system: 
\begin{eqnarray}
%\left\vert \frac{dt}{d\phi} \right\vert 
P > 1.9\,{\rm ms} \, \left\vert \frac{\sin \alpha}{\sin \beta} \right\vert  . 
\end{eqnarray}
This limit can potentially constrain the geometry of the emission. However, the actual rotation period of the system is unknown (under the assumption that the source is rotating): a periodicity search of the burst data revealed no significant detection of a pulse period (see \S~\ref{sec:basichtr}).
We cannot therefore provide an unambiguous interpretation of the above constraint, so we discuss each possibility in turn. The possibility that the entire pulse train is emitted over a single rotation would require $|\sin \alpha/\sin \beta | \lesssim 0.7$, if we approximate $P$ as the total burst duration time $\sim 1.3\,{\rm ms}$.  However, this appears implausible in the context of the rotating-vector model because the position angle of the third pulse is inconsistent with the trend observed in the first pulse.  
The alternate hypothesis is that the pulse train is emitted over multiple rotation periods, in which case the pulse period would either be comparable to the 0.5\,ms duration between each of the four pulses, or the 0.8\,ms duration between pulses one and three and two and four (with each alternate pulse representing interpulse-like emission).  A spin period of 0.5\,ms is implausible since it violates the expected breakup speed for neutron stars \citep{CookShapiroTeukolsky94,HaenselLasotaZdunik99}. A spin period of $P \approx 0.8\,$ms would imply $|\sin \alpha / \sin \beta | < 0.4$, requiring that the magnetic axis is inclined at less than $24^\circ$ to the spin axis in the system.
  
The above argument presupposes rotation of an ordered magnetic field (e.g. one dominated by the dipole component) across the line of sight, and an interpretation of the position angle\,\,in terms of the RVM would be invalid were the magnetic field highly inhomogeneous across the emission region, thus obviating an argument against the four observed pulses not emanating from a single rotation. For instance, millisecond pulsars often exhibit large position-angle gradients with complicated structures that do not lend themselves to a ready interpretation in terms of the RVM (see e.g. \citet{Xilouris1998,Stairs1999,Yan2011}).
Non-linear propagation effects in the pulsar magnetosphere, or perhaps in relativistic plasma surrounding the source, would further complicate this interpretation.
% e.g. Chung and Melatos 2011

%\textcolor{red}{two vector model?}

%\textcolor{red}{comparison with Crab giant pulses?}

%\begin{figure}
%    \plotone{Glensing.pdf}
    %\centering
    %\includegraphics[width=\textwidth]{figures/GalacticLensConstraint.png}
%    \caption{The Poissonian probability of observing lensing (P$_{\text{L}}$) defined in \eqref{eq:LProb} as a function of MACHO mass ($M_L$) for a range of MACHO dark matter fractions (f$_{\text{DM}}$). For FRB\,181112 the cross sections for masses below the black dotted line are magnification limited. Above this cutoff they are time delay limited and heavily mass dependent. The white dotted line indicates the mass where the cross section becomes zero. Above a lens mass of $\sim 1M\odot$, magnification limited cross sections are effectively mass independent.}
%    \label{fig:LConstraints}
%\end{figure}

\subsection{Relativistic Plasma}\label{sec:rel_plasma}

Finally, motivated by the analysis of the total polarization properties of the first pulse, we consider if the polarization properties of the FRB are consistent with propagation through a birefringent region, which could include relativistic or traditional cold plasma or a highly magnetised vacuum. The polarization fraction remaining constant and near 100\% suggests that some of the linear and circular polarization is being interconverted over the duration of the pulse.  
Additionally, the polarization vector of the pulse can be approximately described by a small closed loop on the surface of the \Poincare sphere over the duration of the pulse, as displayed in Figure~\ref{fig:poincare}.

This leads to the possible interpretation that the polarization behavior is a result of propagation through a birefringent plasma; however the conversion between linear and circular polarization requires that the natural modes of the plasma be almost linear (not circular, as in a cold plasma).

In such a medium, the polarization undergoes generalized Faraday rotation \citep[GFR; e.g.][]{KennettMelrose1998}, which causes the polarization vector to rotate at constant latitude about this natural mode axis, with the longitude $2\Psi$ of this polarization vector (with respect to the natural mode axis) scaling  as  $\Psi = \lambda^3\,{\rm RRM}$. ${\rm RRM}$ is a relativistic rotation measure,
%analogous to cold plasma's rotation measure.
defined as ${\rm RRM}=3 \times 10^4\, L \langle n_r B^2 \sin^2 \theta \, \gamma_{\rm min} \rangle $\,rad\,m$^{-3}$, where $L$ is the path length (measured in pc), $n_r$ is the density of relativistic particles (in number per cubic centimeter), $B$ is the magnetic field (in gauss), $\theta$ is the angle between the magnetic field and propagation axis, and $\gamma_{\rm min}$ is the minimum Lorentz factor of the relativistic particle distribution (see \citet{KennettMelrose1998}).
This polarization position angle, $\Psi$, should not be confused with the linear polarization position angle $\chi$.  

The size of the loop on the \Poincare sphere depends on the angle (i.e. the latitude) that the polarization vector makes with the natural mode. 
If the angle is 90$^\circ$, propagation will result in very large changes in the fractional linear and circular polarization.  
If the angle is small, as it appears to be in the present case, the range of values of $Q$ and $U$ probed will also be relatively small.

In this model, the nearly closed loop the polarization vector exhibits over the pulse duration would require a change in RRM over the course of the pulse. For the polarization vector to trace out at least one rotation about the axis of the natural mode we require $\Delta \Psi \sim  \pi \sim \lambda_0^3\,\Delta {\rm RRM}$, where $\lambda_0$ is the central frequency. Thus, a change of at least $\Delta {\rm RRM} \approx 250\,$rad\,m$^{-3}$ is implied over the duration of the pulse.

If this model is correct, then the interpretation of the linear polarization position angle, $\chi$, with frequency in terms of (normal cold plasma) Faraday rotation in Figure \ref{fig:rmdiff} is incorrect. The interconversion is not purely between $Q$ and $U$, but rather $Q$, $U$ and $V$.  Thus, the wrong model is being applied to interpret the linear polarization position angle change as a function of frequency.

If the emitting and birefringent media are spatially separate in the FRB region, then one would expect the polarization longitude to scale as $\lambda^3\,{\rm RRM}$. The fact that we do not see a sign change in $V$ across the band would then place an upper limit on RRM. The change in $\Delta (\lambda^3)$ across the band is $(c/1.13\,{\rm GHz})^3-(c/1.45\,{\rm GHz})^3 \simeq 0.01$, and the requirement that there is no sign change in $V$ at any instant in time across the band implies $\Delta \Psi = \Delta (\lambda^3)\,{\rm RRM} <\pi $, or ${\rm RRM} \lesssim 310\,$rad\,m$^{-3}$.

However, if the birefringent medium and emission region actually are co-located, the frequency dependence of the polarization need not follow a simple $\lambda^3$ dependence and will depend in detail on the geometry of the emission region. An additional potential complication is that, if conditions within the plasma change with time (as is required by the data in the context of this model), the natural modes of the plasma might likewise change on a timescale comparable to the pulse duration. The difference in the frequency dependence of the linear polarization between pulses 1 and 3 (as shown in Figure~\ref{fig:rmdiff}) lends some credence to this hypothesis.

The polarization behavior might instead arise as a result of propagation through a vacuum near the emission region whose magnetic field strength is comparable to the critical magnetic field.  However, the predicted polarization behavior appears inconsistent with this model, for which the natural modes of the region are purely linear. This possibility is still being investigated.
  
%  {\bf Need to refer to model in context of FRB emission models, Belobodorov, Metzger, Maraglit etc.}
%  \cite{Beloborodov2019}
%  \cite{Metzger2019}
  
%There is no evidence for a radio nebula coincident with the FRB. 
There is no evidence for radio synchrotron emission from relativistic plasma coincident with this FRB. This is in contrast to the situation in FRB\,121102 where \citet{Vedantham2019} point out that the relativistic plasma required to produce the observed synchrotron emission does not cause polarization conversion. Given the differences in the properties between FRBs 121102 and 181112, it is unclear of the relevance of bespoke models mooted for the bursts produced by the FRB\,121102 source \citep{Metzger2019,Beloborodov2019}.  

\section{Conclusion}

% RMS:   we don't need to have a long conclusion

Enabled by a new technique that can provide up to $3$-ns time resolution, we have studied the temporal and spectral structure of the ASKAP-localized FRB\,181112 and found the burst can be divided into four pulses.
The pulses show a diversity of phenomenology, which defy a simple explanation, but nevertheless provide strong constraints on both the origin of the emission and multipath propagation along the sightline traversed by the pulse. 

% A short paragraph on what it has stated about the foreground galaxies
Both the time-domain structure of the narrowest pulses (including the lack of a chromatic exponential tail) and the frequency domain structure (autocorrelation function) are consistent with extremely low levels of scattering due to multipath propagation, despite the fact that this radiation passed through an intervening galaxy halo. The limits we obtain on scattering and hence turbulence in the intervening halo are at least a factor of two tighter than those previously reported by \citet{Prochaska+19}, due to the higher time and frequency resolution available to us.
%has allowed us to place better constraints on the propagation through the halo of the its foreground halo. The pulse morphology is inconsistent with arising from scattering in turbulent plasma, so we can constrain the level of turbulence in the plasma.   

% A short paragraph about  possible explanations

% summary of discussion
Several scenarios, summarized in Table \ref{tab:discussion_summary}, are considered in order to explain some of the observed properties of FRB\,181112. 
While the presence of multiple pulses and variation in DM and RM between pulses are consistent with propagation through the foreground halo, the absence of correlation between the pulses and differences in the polarization properties (importantly different polarization position angle swings and circular polarizations) are inconsistent with this scenario.  
In contrast, the properties are more consistent with being produced in the burst source or by propagation through a relativistic plasma, presmably very close to the burst source.
The path traced by the polarization of the first pulse on the \Poincare sphere provides tantalizing evidence that the emission has undergone generalized Faraday rotation.%, but other properties of the pulse cannot be explained by this.

The polarization properties of FRB\,181112 are similar to those shown in some Galactic pulsars and magnetars.
\cite{Illie2019} report apparent rotation measure variations across pulse profiles in a sample of energetic pulsars. They  argue that the results are consistent with distortions due to propagation effects in the neutron star magnetosphere.
Also, the similar polarization conversion between linear and circular is exhibited in the magnetar XTE~J1810$-$197 \citep{Dai2019ApJ} (via private communication) while the high degree of polarization is maintained.

% A short paragraph about the future
% future work
Through an application of the high time resolution voltage reconstruction software to FRB\,181112, we are anticipating to bring a fundamental advance in future studies of FRBs.
This will enable us to study not only the burst's emission mechanism, but also the medium along the propagating path, including relativistic plasma close to the burst source in addition to diffuse matter along the line of sight through plasma and (potentially) gravitational effects.  
We are currently applying the software to all subsequently detected ASKAP FRBs with voltage data products. The results will no doubt identify diversity and commonality in emission and propagation across the fast radio burst population.

\acknowledgments
K.W.B., J.P.M, and R.M.S. acknowledge Australian Research Council (ARC) grant DP180100857.
R.M.S. is the recipient of an ARC Future Fellowship (FT190100155). M.B. and R.M.S. acknowledge support through ARC grant FL150100148. R.M.S. also acknowledges salary support through ARC grant CE170100004. 
A.T.D. is the recipient of an ARC Future Fellowship (FT150100415).
J.X.P. acknowledges support for the F$^4$ collaboration
from NSF grant AST-1911140. This work was performed on the OzSTAR national facility at Swinburne University of Technology. OzSTAR is funded by Swinburne University of Technology and the National Collaborative Research Infrastructure Strategy (NCRIS).
The Australian SKA Pathfinder is part of the Australia Telescope National Facility which is managed by CSIRO. Operation of ASKAP is funded by the Australian Government with support from the National Collaborative Research Infrastructure Strategy. ASKAP uses the resources of the Pawsey Supercomputing Centre. Establishment of ASKAP, the Murchison Radio-astronomy Observatory and the Pawsey Supercomputing Centre are initiatives of the Australian Government, with support from the Government of Western Australia and the Science and Industry Endowment Fund. We acknowledge the Wajarri Yamatji people as the traditional owners of the Observatory site.

%% To help institutions obtain information on the effectiveness of their 
%% telescopes the AAS Journals has created a group of keywords for telescope 
%% facilities.
%
%% Following the acknowledgments section, use the following syntax and the
%% \facility{} or \facilities{} macros to list the keywords of facilities used 
%% in the research for the paper.  Each keyword is check against the master 
%% list during copy editing.  Individual instruments can be provided in 
%% parentheses, after the keyword, but they are not verified.

%\vspace{3mm}
%\facilities{ASKAP?}

%% Similar to \facility{}, there is the optional \software command to allow 
%% authors a place to specify which programs were used during the creation of 
%% the manuscript. Authors should list each code and include either a
%% citation or url to the code inside ()s when available.

\software{PSRCHIVE \citep{Hotan2004},
        DSPSR \citep{vanStraten2011},
        Astropy \citep{2013A&A...558A..33A},
        NumPy \citep{vanderWalt2011NumPy},
        Matplotlib \citep{Hunter2007Matplotlib}
          }

%% Appendix material should be preceded with a single \appendix command.
%% There should be a \section command for each appendix. Mark appendix
%% subsections with the same markup you use in the main body of the paper.

%% Each Appendix (indicated with \section) will be lettered A, B, C, etc.
%% The equation counter will reset when it encounters the \appendix
%% command and will number appendix equations (A1), (A2), etc. The
%% Figure and Table counter will not reset.

%\appendix

%% For this sample we use BibTeX plus aasjournals.bst to generate the
%% the bibliography. The sample63.bib file was populated from ADS. To
%% get the citations to show in the compiled file do the following:
%%
%% pdflatex sample63.tex
%% bibtext sample63
%% pdflatex sample63.tex
%% pdflatex sample63.tex

\vspace{1cm}

\bibliography{frb181112}{}

\newcommand{\noop}[1]{}
\begin{thebibliography}{}
\expandafter\ifx\csname natexlab\endcsname\relax\def\natexlab#1{#1}\fi
\providecommand{\url}[1]{\href{#1}{#1}}
\providecommand{\dodoi}[1]{doi:~\href{http://doi.org/#1}{\nolinkurl{#1}}}
\providecommand{\doeprint}[1]{\href{http://ascl.net/#1}{\nolinkurl{http://ascl.net/#1}}}
\providecommand{\doarXiv}[1]{\href{https://arxiv.org/abs/#1}{\nolinkurl{https://arxiv.org/abs/#1}}}

\bibitem[{{Ashton} {et~al.}(2019){Ashton}, {H{\"u}bner}, {Lasky}, {Talbot},
  {Ackley}, {Biscoveanu}, {Chu}, {Divakarla}, {Easter}, {Goncharov}, {Hernandez
  Vivanco}, {Harms}, {Lower}, {Meadors}, {Melchor}, {Payne}, {Pitkin},
  {Powell}, {Sarin}, {Smith}, \& {Thrane}}]{bilby}
{Ashton}, G., {H{\"u}bner}, M., {Lasky}, P.~D., {et~al.} 2019, \apjs, 241, 27,
  \dodoi{10.3847/1538-4365/ab06fc}

\bibitem[{{Astropy Collaboration} {et~al.}(2013){Astropy Collaboration},
  {Robitaille}, {Tollerud}, {Greenfield}, {Droettboom}, {Bray}, {Aldcroft},
  {Davis}, {Ginsburg}, {Price-Whelan}, {Kerzendorf}, {Conley}, {Crighton},
  {Barbary}, {Muna}, {Ferguson}, {Grollier}, {Parikh}, {Nair}, {Unther},
  {Deil}, {Woillez}, {Conseil}, {Kramer}, {Turner}, {Singer}, {Fox}, {Weaver},
  {Zabalza}, {Edwards}, {Azalee Bostroem}, {Burke}, {Casey}, {Crawford},
  {Dencheva}, {Ely}, {Jenness}, {Labrie}, {Lim}, {Pierfederici}, {Pontzen},
  {Ptak}, {Refsdal}, {Servillat}, \& {Streicher}}]{2013A&A...558A..33A}
{Astropy Collaboration}, {Robitaille}, T.~P., {Tollerud}, E.~J., {et~al.} 2013,
  \aap, 558, A33, \dodoi{10.1051/0004-6361/201322068}

\bibitem[{{Bannister} {et~al.}(2019){Bannister}, {Deller}, {Phillips},
  {Macquart}, {Prochaska}, {Tejos}, {Ryder}, {Sadler}, {Shannon}, {Simha},
  {Day}, {McQuinn}, {North-Hickey}, {Bhandari}, {Arcus}, {Bennert}, {Burchett},
  {Bouwhuis}, {Dodson}, {Ekers}, {Farah}, {Flynn}, {James}, {Kerr}, {Lenc},
  {Mahony}, {O{\textquoteright}Meara}, {Os{\l}owski}, {Qiu}, {Treu}, {U},
  {Bateman}, {Bock}, {Bolton}, {Brown}, {Bunton}, {Chippendale}, {Cooray},
  {Cornwell}, {Gupta}, {Hayman}, {Kesteven}, {Koribalski}, {MacLeod},
  {McClure-Griffiths}, {Neuhold}, {Norris}, {Pilawa}, {Qiao}, {Reynolds},
  {Roxby}, {Shimwell}, {Voronkov}, \& {Wilson}}]{Bannister2019}
{Bannister}, K.~W., {Deller}, A.~T., {Phillips}, C., {et~al.} 2019, Science,
  365, 565, \dodoi{10.1126/science.aaw5903}

\bibitem[{{Bellanger} {et~al.}(1976){Bellanger}, {Bonnerot}, \&
  {Coudreuse}}]{Bellanger1976}
{Bellanger}, M., {Bonnerot}, G., \& {Coudreuse}, M. 1976, IEEE Transactions on
  Acoustics, Speech, and Signal Processing, 24, 109,
  \dodoi{10.1109/TASSP.1976.1162788}

\bibitem[{{Beloborodov}(2019)}]{Beloborodov2019}
{Beloborodov}, A.~M. 2019, arXiv e-prints, arXiv:1908.07743.
\newblock \doarXiv{1908.07743}

\bibitem[{{Blaskiewicz} {et~al.}(1991){Blaskiewicz}, {Cordes}, \&
  {Wasserman}}]{Blaskiewicz91}
{Blaskiewicz}, M., {Cordes}, J.~M., \& {Wasserman}, I. 1991, \apj, 370, 643,
  \dodoi{10.1086/169850}

\bibitem[{{Chashei} \& {Shishov}(1976)}]{Chashei1976_ACFfit}
{Chashei}, I.~V., \& {Shishov}, V.~I. 1976, \sovast, 20, 13

\bibitem[{{Chatterjee} {et~al.}(2017){Chatterjee}, {Law}, {Wharton},
  {Burke-Spolaor}, {Hessels}, {Bower}, {Cordes}, {Tendulkar}, {Bassa},
  {Demorest}, {Butler}, {Seymour}, {Scholz}, {Abruzzo}, {Bogdanov}, {Kaspi},
  {Keimpema}, {Lazio}, {Marcote}, {McLaughlin}, {Paragi}, {Ransom}, {Rupen},
  {Spitler}, \& {van Langevelde}}]{Chatterjee2017}
{Chatterjee}, S., {Law}, C.~J., {Wharton}, R.~S., {et~al.} 2017, \nat, 541, 58,
  \dodoi{10.1038/nature20797}

\bibitem[{{Clarke} {et~al.}(2014){Clarke}, {D'Addario}, {Navarro}, \&
  {Trinh}}]{Clarke2014}
{Clarke}, N., {D'Addario}, L., {Navarro}, R., \& {Trinh}, J. 2014, Journal of
  Astronomical Instrumentation, 3, 1450004, \dodoi{10.1142/S2251171714500044}

\bibitem[{{Cook} {et~al.}(1994){Cook}, {Shapiro}, \&
  {Teukolsky}}]{CookShapiroTeukolsky94}
{Cook}, G.~B., {Shapiro}, S.~L., \& {Teukolsky}, S.~A. 1994, \apj, 424, 823,
  \dodoi{10.1086/173934}

\bibitem[{{Cordes}(1986)}]{Cordes1986ApJ_ACF}
{Cordes}, J.~M. 1986, \apj, 311, 183, \dodoi{10.1086/164764}

\bibitem[{{Cordes} {et~al.}(2017){Cordes}, {Wasserman}, {Hessels}, {Lazio},
  {Chatterjee}, \& {Wharton}}]{Cordes2017}
{Cordes}, J.~M., {Wasserman}, I., {Hessels}, J.~W.~T., {et~al.} 2017, \apj,
  842, 35, \dodoi{10.3847/1538-4357/aa74da}

\bibitem[{{Cordes} {et~al.}(1983){Cordes}, {Weisberg}, \&
  {Boriakoff}}]{Cordes1983ApJ_ACF}
{Cordes}, J.~M., {Weisberg}, J.~M., \& {Boriakoff}, V. 1983, \apj, 268, 370,
  \dodoi{10.1086/160961}

\bibitem[{{Dai} {et~al.}(2019){Dai}, {Lower}, {Bailes}, {Camilo}, {Halpern},
  {Johnston}, {Kerr}, {Reynolds}, {Sarkissian}, \& {Scholz}}]{Dai2019ApJ}
{Dai}, S., {Lower}, M.~E., {Bailes}, M., {et~al.} 2019, \apjl, 874, L14,
  \dodoi{10.3847/2041-8213/ab0e7a}

\bibitem[{{Farah} {et~al.}(2018){Farah}, {Flynn}, {Bailes}, {Jameson},
  {Bannister}, {Barr}, {Bateman}, {Bhand ari}, {Caleb}, {Campbell-Wilson},
  {Chang}, {Deller}, {Green}, {Hunstead}, {Jankowski}, {Keane}, {Macquart},
  {M{\"o}ller}, {Onken}, {Os{\l}owski}, {Parthasarathy}, {Plant}, {Ravi},
  {Shannon}, {Tucker}, {Venkatraman Krishnan}, \& {Wolf}}]{Farah2018}
{Farah}, W., {Flynn}, C., {Bailes}, M., {et~al.} 2018, \mnras, 478, 1209,
  \dodoi{10.1093/mnras/sty1122}

\bibitem[{{Farah} {et~al.}(2019){Farah}, {Flynn}, {Bailes}, {Jameson},
  {Bateman}, {Campbell-Wilson}, {Day}, {Deller}, {Green}, {Gupta}, {Hunstead},
  {Lower}, {Os{\l}owski}, {Parthasarathy}, {Price}, {Ravi}, {Shannon},
  {Sutherland }, {Temby}, {Krishnan}, {Caleb}, {Chang}, {Cruces}, {Roy},
  {Morello}, {Onken}, {Stappers}, {Webb}, \& {Wolf}}]{Farah2019}
---. 2019, \mnras, 488, 2989, \dodoi{10.1093/mnras/stz1748}

\bibitem[{{Haensel} {et~al.}(1999){Haensel}, {Lasota}, \&
  {Zdunik}}]{HaenselLasotaZdunik99}
{Haensel}, P., {Lasota}, J.~P., \& {Zdunik}, J.~L. 1999, \aap, 344, 151

\bibitem[{{Hankins}(1971)}]{Hankins1971}
{Hankins}, T.~H. 1971, \apj, 169, 487, \dodoi{10.1086/151164}

\bibitem[{{Hankins}(1972)}]{Hankins1972}
---. 1972, \apjl, 177, L11, \dodoi{10.1086/181043}

\bibitem[{{Hankins} \& {Rickett}(1975)}]{Hankins1975}
{Hankins}, T.~H., \& {Rickett}, B.~J. 1975, Methods in Computational Physics,
  14, 55

\bibitem[{{Hessels} {et~al.}(2019){Hessels}, {Spitler}, {Seymour}, {Cordes},
  {Michilli}, {Lynch}, {Gourdji}, {Archibald}, {Bassa}, {Bower}, {Chatterjee},
  {Connor}, {Crawford}, {Deneva}, {Gajjar}, {Kaspi}, {Keimpema}, {Law},
  {Marcote}, {McLaughlin}, {Paragi}, {Petroff}, {Ransom}, {Scholz}, {Stappers},
  \& {Tendulkar}}]{Hessels2019}
{Hessels}, J.~W.~T., {Spitler}, L.~G., {Seymour}, A.~D., {et~al.} 2019, \apjl,
  876, L23, \dodoi{10.3847/2041-8213/ab13ae}

\bibitem[{{Hibschman} \& {Arons}(2001)}]{HibschmanArons01}
{Hibschman}, J.~A., \& {Arons}, J. 2001, \apj, 546, 382, \dodoi{10.1086/318224}

\bibitem[{{Hotan} {et~al.}(2004){Hotan}, {van Straten}, \&
  {Manchester}}]{Hotan2004}
{Hotan}, A.~W., {van Straten}, W., \& {Manchester}, R.~N. 2004, \pasa, 21, 302,
  \dodoi{10.1071/AS04022}

\bibitem[{{Hunter}(2007)}]{Hunter2007Matplotlib}
{Hunter}, J.~D. 2007, Computing in Science and Engineering, 9, 90,
  \dodoi{10.1109/MCSE.2007.55}

\bibitem[{{Ilie} {et~al.}(2019){Ilie}, {Johnston}, \& {Weltevrede}}]{Illie2019}
{Ilie}, C.~D., {Johnston}, S., \& {Weltevrede}, P. 2019, \mnras, 483, 2778,
  \dodoi{10.1093/mnras/sty3315}

\bibitem[{{Kennett} \& {Melrose}(1998)}]{KennettMelrose1998}
{Kennett}, M., \& {Melrose}, D. 1998, \pasa, 15, 211, \dodoi{10.1071/AS98211}

\bibitem[{{Lambert} \& {Rickett}(1999)}]{Lambert1999}
{Lambert}, H.~C., \& {Rickett}, B.~J. 1999, \apj, 517, 299,
  \dodoi{10.1086/307181}

\bibitem[{{Lange} {et~al.}(1998){Lange}, {Kramer}, {Wielebinski}, \&
  {Jessner}}]{Lange1998}
{Lange}, C., {Kramer}, M., {Wielebinski}, R., \& {Jessner}, A. 1998, \aap, 332,
  111

\bibitem[{{Lorimer} \& {Kramer}(2012)}]{Lormier2012book}
{Lorimer}, D.~R., \& {Kramer}, M. 2012, {Handbook of Pulsar Astronomy}

\bibitem[{{Lu} \& {Phinney}(2019)}]{Lu2019}
{Lu}, W., \& {Phinney}, E.~S. 2019, arXiv e-prints, arXiv:1912.12241.
\newblock \doarXiv{1912.12241}

\bibitem[{{Lyutikov}(2016)}]{Lyutikov16}
{Lyutikov}, M. 2016, arXiv e-prints, arXiv:1607.00777.
\newblock \doarXiv{1607.00777}

\bibitem[{{Lyutikov}(2019{\natexlab{a}})}]{Lyutikov19b}
---. 2019{\natexlab{a}}, arXiv e-prints, arXiv:1909.10409.
\newblock \doarXiv{1909.10409}

\bibitem[{{Lyutikov}(2019{\natexlab{b}})}]{Lyutikov19a}
---. 2019{\natexlab{b}}, arXiv e-prints, arXiv:1901.03260.
\newblock \doarXiv{1901.03260}

\bibitem[{{Macquart} {et~al.}(2010){Macquart}, {Bailes}, {Bhat}, {Bower},
  {Bunton}, {Chatterjee}, {Colegate}, {Cordes}, {D'Addario}, {Deller},
  {Dodson}, {Fender}, {Haines}, {Halll}, {Harris}, {Hotan}, {Johnston},
  {Jones}, {Keith}, {Koay}, {Lazio}, {Majid}, {Murphy}, {Navarro}, {Phillips},
  {Quinn}, {Preston}, {Stansby}, {Stairs}, {Stappers}, {Staveley-Smith},
  {Tingay}, {Thompson}, {van Straten}, {Wagstaff}, {Warren}, {Wayth}, {Wen}, \&
  {CRAFT Collaboration}}]{Macquart2010}
{Macquart}, J.-P., {Bailes}, M., {Bhat}, N.~D.~R., {et~al.} 2010, \pasa, 27,
  272, \dodoi{10.1071/AS09082}

\bibitem[{{Metzger} {et~al.}(2019){Metzger}, {Margalit}, \&
  {Sironi}}]{Metzger2019}
{Metzger}, B.~D., {Margalit}, B., \& {Sironi}, L. 2019, \mnras, 485, 4091,
  \dodoi{10.1093/mnras/stz700}

\bibitem[{{Morrison} {et~al.}(2019){Morrison}, {Bunton}, {van Straten},
  {Deller}, \& {Jameson}}]{Morrison2019}
{Morrison}, I., {Bunton}, J., {van Straten}, W., {Deller}, A.~T., \& {Jameson},
  A. 2019, Journal of Astronomical Instrumentation, accepted

\bibitem[{{Platts} {et~al.}(2018){Platts}, {Weltman}, {Walters}, {Tendulkar},
  {Gordin}, \& {Kandhai}}]{FRBtheoryCat}
{Platts}, E., {Weltman}, A., {Walters}, A., {et~al.} 2018, arXiv e-prints.
\newblock \doarXiv{1810.05836}

\bibitem[{Princen \& Bradley(1986)}]{Princen1986}
Princen, J.~P., \& Bradley, A.~B. 1986, {IEEE} Trans. Acoustics, Speech, and
  Signal Processing, 34, 1153

\bibitem[{{Prochaska} {et~al.}(2019){Prochaska}, {Macquart}, {McQuinn},
  {Simha}, {Shannon}, {Day}, {Marnoch}, {Ryder}, {Deller}, {Bannister},
  {Bhandari}, {Bordoloi}, {Bunton}, {Cho}, {Flynn}, {Mahony}, {Phillips},
  {Qiu}, \& {Tejos}}]{Prochaska+19}
{Prochaska}, J.~X., {Macquart}, J.-P., {McQuinn}, M., {et~al.} 2019, Science,
  365, 231, \dodoi{10.1126/science.aay0073}

\bibitem[{{Radhakrishnan} \& {Cooke}(1969)}]{RadhakrishnanCooke1969}
{Radhakrishnan}, V., \& {Cooke}, D.~J. 1969, \aplett, 3, 225

\bibitem[{{Ransom}(2001)}]{Ransom2001_PRESTO}
{Ransom}, S.~M. 2001, PhD thesis, Harvard University

\bibitem[{{Ravi} {et~al.}(2019){Ravi}, {Catha}, {D'Addario}, {Djorgovski},
  {Hallinan}, {Hobbs}, {Kocz}, {Kulkarni}, {Shi}, {Vedantham}, {Weinreb}, \&
  {Woody}}]{Ravi2019_localisation}
{Ravi}, V., {Catha}, M., {D'Addario}, L., {et~al.} 2019, \nat, 572, 352,
  \dodoi{10.1038/s41586-019-1389-7}

\bibitem[{{Shannon} {et~al.}(2018){Shannon}, {Macquart}, {Bannister}, {Ekers},
  {James}, {Os{\l}owski}, {Qiu}, {Sammons}, {Hotan}, {Voronkov}, {Beresford},
  {Brothers}, {Brown}, {Bunton}, {Chippendale}, {Haskins}, {Leach},
  {Marquarding}, {McConnell}, {Pilawa}, {Sadler}, {Troup}, {Tuthill},
  {Whiting}, {Allison}, {Anderson}, {Bell}, {Collier}, {G{\"u}rkan}, {Heald},
  \& {Riseley}}]{Shannon2018}
{Shannon}, R.~M., {Macquart}, J.-P., {Bannister}, K.~W., {et~al.} 2018, \nat,
  562, 386, \dodoi{10.1038/s41586-018-0588-y}

\bibitem[{{Speagle}(2019)}]{dynesty}
{Speagle}, J.~S. 2019, arXiv e-prints, arXiv:1904.02180.
\newblock \doarXiv{1904.02180}

\bibitem[{{Stairs} {et~al.}(1999){Stairs}, {Thorsett}, \&
  {Camilo}}]{Stairs1999}
{Stairs}, I.~H., {Thorsett}, S.~E., \& {Camilo}, F. 1999, \apjs, 123, 627,
  \dodoi{10.1086/313245}

\bibitem[{{van der Walt} {et~al.}(2011){van der Walt}, {Colbert}, \&
  {Varoquaux}}]{vanderWalt2011NumPy}
{van der Walt}, S., {Colbert}, S.~C., \& {Varoquaux}, G. 2011, Computing in
  Science and Engineering, 13, 22, \dodoi{10.1109/MCSE.2011.37}

\bibitem[{{van Straten} \& {Bailes}(2011)}]{vanStraten2011}
{van Straten}, W., \& {Bailes}, M. 2011, \pasa, 28, 1, \dodoi{10.1071/AS10021}

\bibitem[{{van Straten} {et~al.}(2010){van Straten}, {Manchester}, {Johnston},
  \& {Reynolds}}]{vanStraten2010}
{van Straten}, W., {Manchester}, R.~N., {Johnston}, S., \& {Reynolds}, J.~E.
  2010, \pasa, 27, 104, \dodoi{10.1071/AS09084}

\bibitem[{{Vedantham} \& {Ravi}(2019)}]{Vedantham2019}
{Vedantham}, H.~K., \& {Ravi}, V. 2019, \mnras, 485, L78,
  \dodoi{10.1093/mnrasl/slz038}

\bibitem[{Vetterli \& Kova\v{c}evic(1995)}]{Vetterli1995:WSC}
Vetterli, M., \& Kova\v{c}evic, J. 1995, Wavelets and Subband Coding (Upper
  Saddle River, NJ, USA: Prentice-Hall, Inc.)

\bibitem[{{Xilouris} {et~al.}(1998){Xilouris}, {Kramer}, {Jessner}, {von
  Hoensbroech}, {Lorimer}, {Wielebinski}, {Wolszczan}, \&
  {Camilo}}]{Xilouris1998}
{Xilouris}, K.~M., {Kramer}, M., {Jessner}, A., {et~al.} 1998, \apj, 501, 286,
  \dodoi{10.1086/305791}

\bibitem[{{Yan} {et~al.}(2011){Yan}, {Manchester}, {van Straten}, {Reynolds},
  {Hobbs}, {Wang}, {Bailes}, {Bhat}, {Burke-Spolaor}, {Champion}, {Coles},
  {Hotan}, {Khoo}, {Oslowski}, {Sarkissian}, {Verbiest}, \&
  {Yardley}}]{Yan2011}
{Yan}, W.~M., {Manchester}, R.~N., {van Straten}, W., {et~al.} 2011, \mnras,
  414, 2087, \dodoi{10.1111/j.1365-2966.2011.18522.x}

\end{thebibliography}
\bibliographystyle{aasjournal}

%% This command is needed to show the entire author+affiliation list when
%% the collaboration and author truncation commands are used.  It has to
%% go at the end of the manuscript.
%\allauthors

%% Include this line if you are using the \added, \replaced, \deleted
%% commands to see a summary list of all changes at the end of the article.
%\listofchanges

\end{document}